\documentclass[twocolumn, tighten]{aastex631}

\providecommand{\dt}[1]{{\tt #1}}
\usepackage{amsmath}
\usepackage{comment}
\usepackage{threeparttable}
\usepackage{multirow}

\usepackage{CJKutf8}

\begin{document}

\title{Identifying the Galactic Substructures in 5D Space Using All-sky RR Lyrae Stars in Gaia DR3}

\correspondingauthor{Huawei Zhang, Xiang-Xiang Xue, Yang Huang}
\email{zhanghw@pku.edu.cn, xuexx@nao.cas.cn, huangyang@ucas.ac.cn}
\newcommand{\KIAA}{\affiliation{Kavli Institute for Astronomy and
Astrophysics, Peking University, Beijing 100871, China}}
\newcommand{\DoA}{\affiliation{Department of Astronomy, School of Physics,
Peking University, Beijing 100871, China}}
\newcommand{\UCAS}{\affiliation{School of Astronomy and Space Science, University of Chinese Academy of Sciences, Beijing 100049, China}}
\newcommand{\NAOC}{\affiliation{CAS Key Laboratory of Optical Astronomy, National Astronomical Observatories, Chinese Academy of Sciences, Beijing 100101, China}}
\newcommand{\IFAA}{\affiliation{Institute for Frontiers in Astronomy and Astrophysics, Beijing Normal University, Beijing 102206, China}}
\newcommand{\MPI}{\affiliation{Max-Planck-Institute for Astronomy K\"{o}nigstuhl 17, D-69117, Heidelberg, Germany}}
\newcommand{\QNU}{\affiliation{College of Physics and Electronic Engineering, Qilu Normal University, Jinan 250200, China}}
\newcommand{\TGU}{\affiliation{Center for Astronomy and Space Sciences, China Three Gorges University, Yichang 443002, China}}
\newcommand{\SHAO}{\affiliation{Shanghai Astronomical Observatory, 80 Nandan Road, Shanghai 200030, China}}
\newcommand{\THU}{\affiliation{Department of Astronomy, Tsinghua University, Beijing 100084, China}}

\author[0000-0002-9796-1507]{Shenglan Sun}\DoA\KIAA
\author[0000-0001-6580-1552]{Fei Wang}\DoA\KIAA
\author[0000-0002-7727-1699]{Huawei Zhang}\DoA\KIAA
\author[0000-0002-0642-5689]{Xiang-Xiang Xue}\NAOC\IFAA
\author[0000-0003-3250-2876]{Yang Huang}\UCAS\NAOC
\author[0009-0008-1319-1084]{Ruizhi Zhang}\NAOC\UCAS
\author[0000-0003-4996-9069]{Hans-Walter Rix}\MPI
\author[0000-0003-2086-0684]{Xinyi Li}\QNU
\author{Gaochao Liu}\TGU
\author[0000-0001-7080-0618]{Lan Zhang}\NAOC
\author[0000-0003-1972-0086]{Chengqun Yang}\SHAO
\author[0000-0003-1454-1636]{Shuo Zhang}\THU

%% Note that the \and command from previous versions of AASTeX is now
%% depreciated in this version as it is no longer necessary. AASTeX 
%% automatically takes care of all commas and "and"s between authors names.

%% AASTeX 6.31 has the new \collaboration and \nocollaboration commands to
%% provide the collaboration status of a group of authors. These commands 
%% can be used either before or after the list of corresponding authors. The
%% argument for \collaboration is the collaboration identifier. Authors are
%% encouraged to surround collaboration identifiers with ()s. The 
%% \nocollaboration command takes no argument and exists to indicate that
%% the nearby authors are not part of surrounding collaborations.

%% Mark off the abstract in the ``abstract'' environment. 
\begin{abstract}

Motivated by the vast gap between photometric and spectroscopic data volumes, there is great potential in using 5D kinematic information to identify and study substructures of the Milky Way. We identify substructures in the Galactic halo using 46,575 RR Lyrae stars (RRLs) from Gaia DR3 with the photometric metallicities and distances newly estimated by \citet{2023ApJ...944...88L}. Assuming a Gaussian prior distribution of radial velocity, we calculate the orbital distribution characterized by the integrals of motion for each RRL based on its 3D positions, proper motions and corresponding errors, and then apply the friends-of-friends algorithm to identify groups moving along similar orbits. We have identified several known substructures, including Sagittarius (Sgr) Stream, Hercules-Aquila Cloud (HAC), Virgo Overdensity (VOD), Gaia-Enceladus-Sausage (GES), Orphan-Chenab stream, Cetus-Palca, Helmi Streams, Sequoia, Wukong and Large Magellanic Cloud (LMC) leading arm, along with 18 unknown groups. Our findings indicate that HAC and VOD have kinematic and chemical properties remarkably similar to GES, with most HAC and VOD members exhibiting eccentricity as high as GES, suggesting that they may share a common origin with GES. The ability to identify the low mass and spatially dispersed substructures further demonstrates the potential of our method, which breaks the limit of spectroscopic survey and is competent to probe the substructures in the whole Galaxy. Finally, we have also identified 18 unknown groups with good spatial clustering and proper motion consistency, suggesting more excavation of Milky Way substructures in the future with only 5D data.

\end{abstract}

%% Keywords should appear after the \end{abstract} command. 
%% The AAS Journals now uses Unified Astronomy Thesaurus concepts:
%% https://astrothesaurus.org
%% You will be asked to selected these concepts during the submission process
%% but this old "keyword" functionality is maintained in case authors want
%% to include these concepts in their preprints.

%% From the front matter, we move on to the body of the paper.
%% Sections are demarcated by \section and \subsection, respectively.
%% Observe the use of the LaTeX \label
%% command after the \subsection to give a symbolic KEY to the
%% subsection for cross-referencing in a \ref command.
%% You can use LaTeX's \ref and \label commands to keep track of
%% cross-references to sections, equations, tables, and figures.
%% That way, if you change the order of any elements, LaTeX will
%% automatically renumber them.
%%
%% We recommend that authors also use the natbib \citep
%% and \citet commands to identify citations.  The citations are
%% tied to the reference list via symbolic KEYs. The KEY corresponds
%% to the KEY in the \bibitem in the reference list below. 

\section{Introduction}
The $\rm {\Lambda}CDM$ cosmological model predicts that the large-scale structures form through hierarchical processes \citep{1974ApJ...189L..51P,1978MNRAS.183..341W,1984Natur.311..517B}. This model suggests that the Milky Way (MW) developed through a series of accretion and merger events \citep[e.g.,][]{1978ApJ...225..357S,1978MNRAS.183..341W,1984Natur.311..517B}, leading to the tidal disruption of accreted satellite galaxies and the formation of substructures in the stellar halo \citep[e.g.,][]{2001ApJ...548...33B,2005ApJ...635..931B,2010MNRAS.406..744C}. These substructures could remain coherent in phase space for many gigayears \citep{1996ApJ...465..278J,1999MNRAS.307..495H}, providing valuable insights into the formation history of the MW. Chemical properties also serve as key diagnostics for the origins of these substructures \citep{2002ARA&A..40..487F}.

In recent years, the combination of Gaia \citep{2016A&A...595A...1G,2018A&A...616A...1G,2021A&A...649A...1G,2023A&A...674A...1G} and spectroscopic surveys such as LAMOST \citep{2012RAA....12.1197C,2012RAA....12..735D,2012RAA....12..723Z}, APOGEE \citep{2017AJ....154...94M}, SDSS/SEGUE \citep{2000AJ....120.1579Y,2009AJ....137.4377Y} and GALAH \citep{2015GALAH}, has contributed to significant progress in mapping the substructures of the local Galactic halo using full 6D kinematic information \citep[e.g.,][]{2018Natur.563...85H,2018MNRAS.478..611B,2019ApJ...880...65Y,2019A&A...625A...5K,2020ApJ...901...48N,2022MNRAS.513.1958W,2023MNRAS.tmp.3593G,2023MNRAS.520.5225M,zhang2024VMP}. One of the most exciting pieces of the mosaic of Milky Way's evolution history is Gaia-Enceladus-Sausage (GES), which is the last major merger that the Galaxy experienced around 10 Gyr ago \citep{2018Natur.563...85H,2018MNRAS.478..611B}, constituting the majority of the halo stars in the inner halo. Several smaller but also significant substructures including Wukong/LMS-1, Arjuna, Sequoia and I'itoi have been suggested to be part of the GES \citep[e.g.,][]{2020ApJ...898L..37Y,2020ApJ...901...48N,2023MNRAS.520.5671H}. The members of each kinematically coherent substructure also show common chemical abundance patterns, indicating their sharing original accretion events. In a word, we are experiencing a golden era that the local halo substructures have been explored nearly to its fullest with complete kinematic information provided by astrometry and radial velocity measurements from spectroscopy, moving to the whole view of the Galaxy's past.

However, the current sizes and distances of 6D kinematic samples are insufficient for probing the substructures in the outer halo. The sizes of the samples with full 6D information in previous studies \citep{2022MNRAS.513.1958W,2019ApJ...880...65Y,2022ApJ...926...26S,2023MNRAS.tmp.3593G,2020ApJ...901...48N} were limited to $\sim10^3$ to $\sim10^4$, which were also sparse beyond $20\,\text{kpc}$ due to shallower limiting magnitudes of spectroscopic surveys. To investigate the Galactic history thoroughly, we must climb towards a deeper set of questions: On scales comparable to MW's virial radius, how many progenitor galaxies have merged with our Galaxy? What are the composition ratios of different substructures? What are the clear correspondences between various substructures and distinct accretion events? Using larger and further halo star samples are crucial to unravel the answers.

The current limitation of the size of full 6D sample is due to the enormous gap between the amount of spectroscopic data and astrometric/photometric data, with the former being two orders of magnitude smaller than the latter. In Gaia DR3 \citep{2023A&A...674A...1G}, over one billion stars have parallax and proper motion measurements and only 33 million targets have radial velocity measurements in the Radial Velocity Spectrometer (RVS) catalog, which also means that over 97\% of stars only have at most 5D kinematic information and lack radial velocities. All the spectral data until now add up to only tens of millions, of which LAMOST has contributed the vast majority. 

Therefore, it is meaningful to consider the possibility and advantages of studying substructures using only 5D kinematic information, excluding radial velocities. Xue-X.X. et al. (2024, in prep.) proposes that it is efficient to identify substructures in the integrals of motion (IoM) space instead of in the position-velocity space and validates that it is also efficient to identify substructures in the IoM space by using a prior when stars have incomplete velocity information, e.g., lack of proper motions or radial velocities. So there is an urgent need to develop methods to identify substructures using only 5D information. Such methods also have a wide application prospect for the numerous samples from future deep-field photometric surveys such as Large Synoptic Survey Telescope \citep[LSST;][]{2019ApJ...873..111I} and the China Space Station Telescope \citep[CSST;][]{2011SSPMA..41.1441Z,2018MNRAS.480.2178C,2019ApJ...883..203G}. In addition, studying halo substructures using 5D kinematic samples without radial velocities helps to reach more complete sky coverage, avoiding selection effects of sky regions. The previous complete 6D samples were limited by the sky coverage of spectroscopic surveys, potentially introducing selection effects and providing an incomplete understanding of the overall structure of the substructures.

For halo stars farther than $\sim10\,\rm kpc$, the large parallax uncertainties render distance estimates unreliable. Stars with precise distance estimates, such as K giants, blue horizontal-branch (BHB), and RRL stars, are ideal tracers for studying Galactic substructures \citep{2019ApJ...880...65Y,2020ApJ...898L..37Y,2020ARA&A..58..205H,2022MNRAS.513.1958W}. Combined with Gaia's precise celestial positions and proper motions, these stars serve as ideal tracers for implementing substructure identification methods with 5D information.

RRL variables, with their robust luminosity-metallicity relation in the optical band and period-absolute magnitude-metallicity ($PMZ$) relation in the infrared band, are particularly effective for such studies \citep{2008ApJ...678..851K,2013AJ....146...21S,2017ApJ...844L...4S}. \citet{2023MNRAS.tmp.3593G} used 5,355 RRLs selected from the RRL catalog of \citet{2023ApJ...944...88L} with radial velocities mainly from Gaia DR3 \citep{2023A&A...674A...1G} to identify 97 Dynamically Tagged Groups (DTGs), yielding important constraints on halo substructures.

In this work, we aim to identify Galactic halo substructures in the IoM space using a prior distribution of radial velocities as the first attempt of the method using 5D information. We employ large RRL catalogues with precise photometric metallicity and distance estimates from \citet{2023ApJ...944...88L} and proper motions from Gaia DR3 and extend sample size to $\sim10^5$. It is also worth noting that this sample is the first all-sky sample to study halo substructures.

This paper is organized as follows. The data employed in this work is described in Section~\ref{sec:style}. In Section~\ref{sec:method}, we describe the group identification approach and validate our method. We present the results in Section~\ref{sec:results}. Finally, a summary concludes the paper in Section~\ref{sec:summary}.

\section{Data} \label{sec:style}

\begin{figure}
	\centering
	\includegraphics[width = \linewidth]{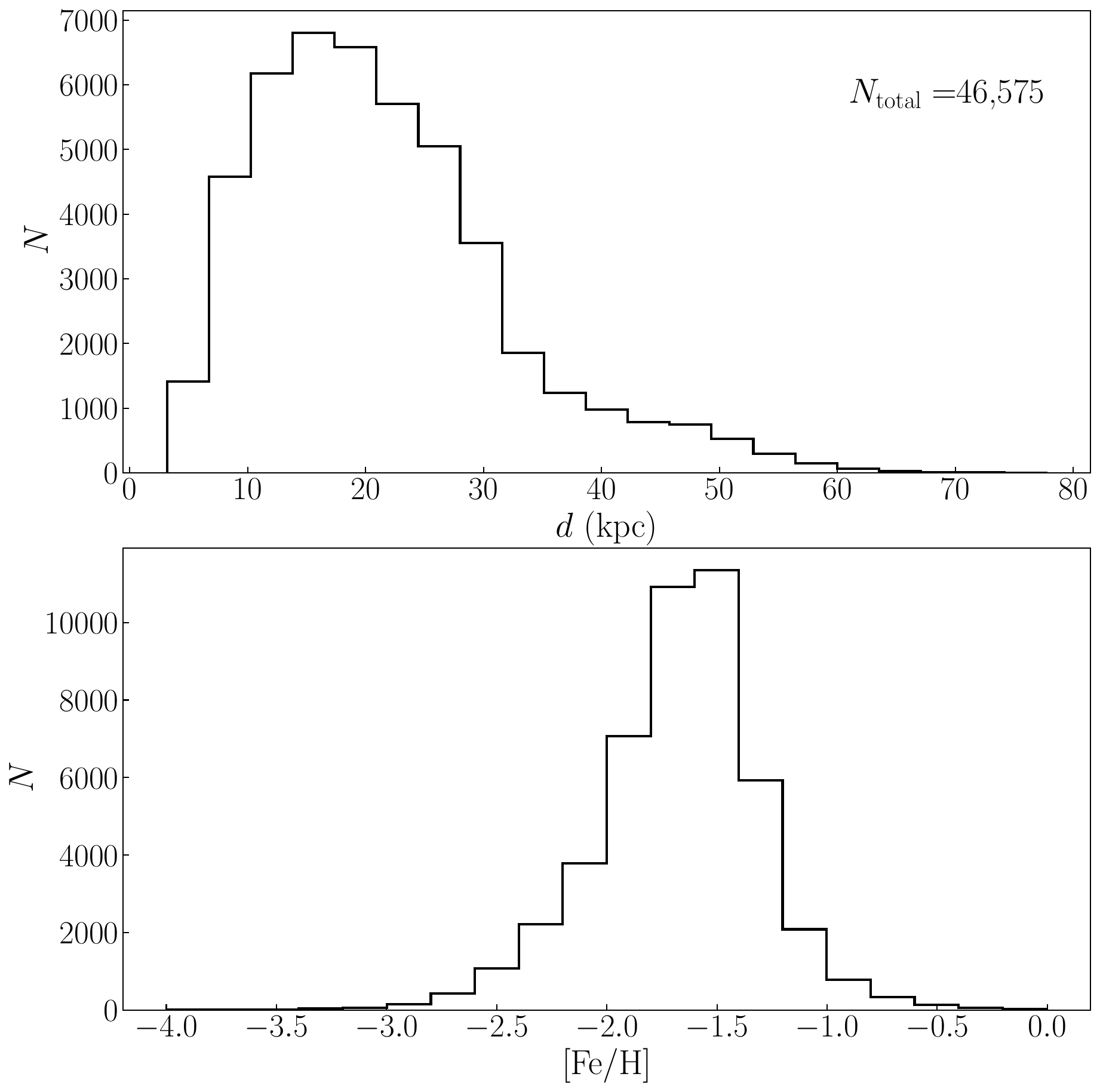}
	\caption{The heliocentric distances and the metallicity distributions of our sample.}
	\centering
	\label{fig:allsample_d_FeH}
\end{figure}

\begin{figure*}
	\centering
	\includegraphics[width = \linewidth]{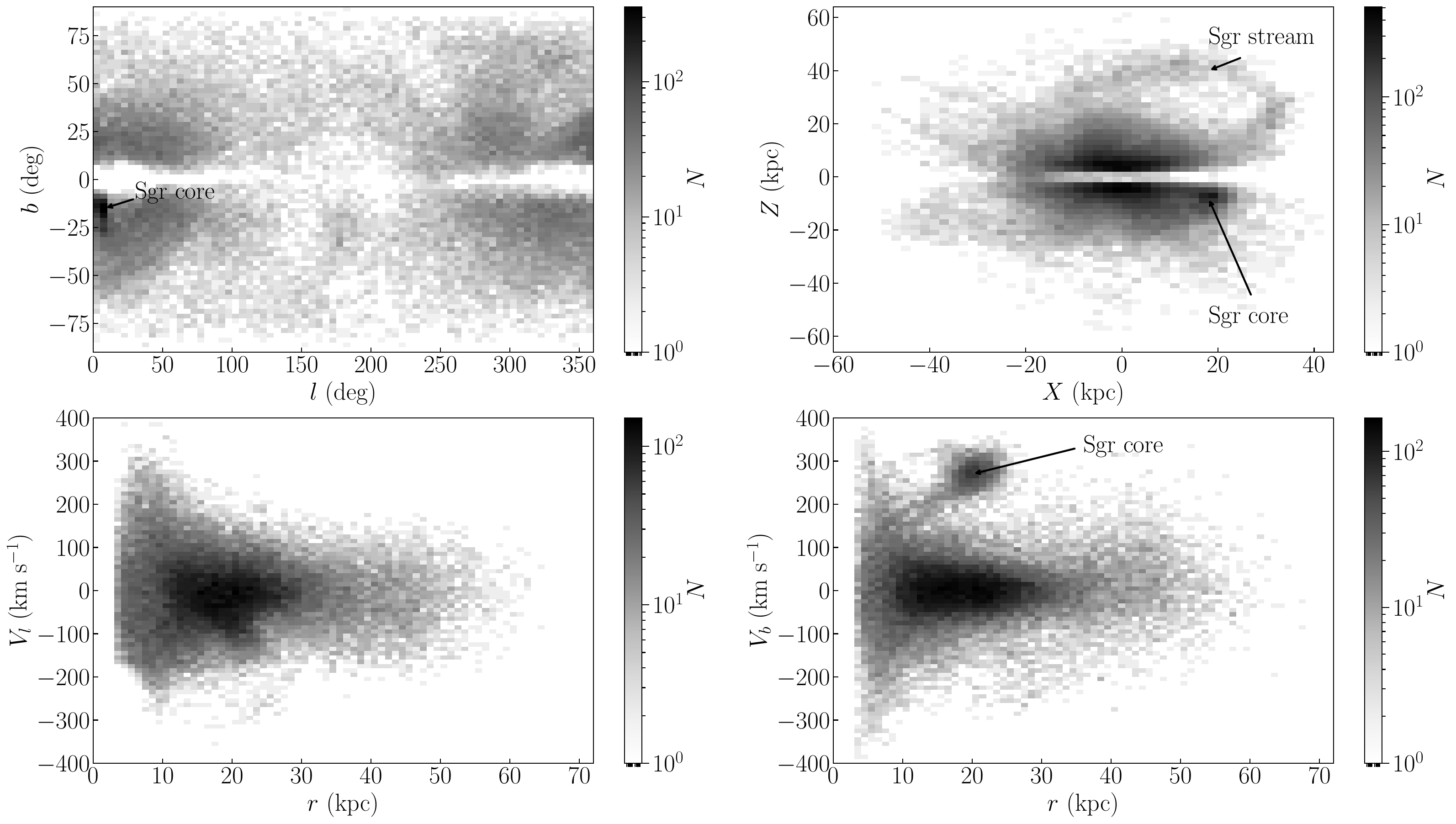}
	\caption{The density map of our RRL sample in the $(l,b)$, $(X,Z)$, $(r,V_l)$, and $(r,V_b)$ spaces, respectively. The Sgr core and Sgr tidal stream are evident in the spaces.}
	\centering
	\label{fig:all_sample_dens}
\end{figure*}

\subsection{RRL sample}\label{subsec:RRLsample}
In this work, we use the catalog of \citet{2023ApJ...944...88L}, which provides estimates of photometric metallicities and distances for 135,873 RRLs (115,410 type RRab and 20,463 type RRc stars) based on the Gaia DR3 RRL catalogue \citep{2023A&A...674A..18C}. \citet{2023A&A...674A..18C} provides precise measurements of parameters derived from Gaia DR3 light curves such as the period $P$ and Fourier-decomposition parameters $\phi_{31}$ and $R_{21}$. Taking advantage of the large RRL catalogue with spectroscopic metallicities from \citet{2020ApJS..247...68L}, \citet{2023ApJ...944...88L} refined the $P-\phi_{31}-R_{21}-[\mathrm{Fe} / \mathrm{H}]$ and $P-R_{21}-[\mathrm{Fe} / \mathrm{H}]$ relations for type RRab and type RRc stars, achieving re-estimated photometric metallicity uncertainties of 0.24 dex for RRab and $0.16\,\rm dex$ for RRc, respectively. As for distances, they re-calibrated the $M_G-{\rm [Fe/H]}$, $PM_{K_s}Z$ and $PM_{W1}Z$ relations of RRLs and yielded the distances after correcting the interstellar extinction. $M_G$, $M_{K_s}$ and $M_{W1}$ are the absolute magnitude of $G$, near-infrared $K_s$ and mid-infrared $W1$ passbands for Gaia, 2MASS, and the Wide-field Infrared Survey Explorer \citep[WISE;][]{2010AJ....140.1868W}. The typical uncertainties of the distances are 9.91\% for RRab and 9.77\% for RRc. To eliminate the unreliable data, we select the RRab and RRc with $\rm -4<[Fe/H]<0$ and remove stars with the metallicity uncertainty larger than $1.0\,\rm dex$ and the ones with the distance uncertainty larger than 20\%. 

We clean the RRL sample by removing the RRLs associated with known structures, the Magellanic Clouds and globular clusters (GCs), to focus on the substructures in the Milky Way. We adopt the centers of the Large Magellanic Cloud (LMC) and Small Magellanic Cloud (SMC) at $(\alpha_{\rm LMC},\delta_{\rm LMC})=(81^{\circ}.28,-69^{\circ}.78)$ \citep{2001AJ....122.1827V} and $(\alpha_{\rm SMC},\delta_{\rm SMC})=(12^{\circ}.80,-73^{\circ}.15)$ \citep{2000A&A...358L...9C}. The distances of LMC and SMC we adopt are $49.89\,\rm kpc$ \citep{2014AJ....147..122D} and $61.94\,\rm kpc$ \citep{2015AJ....149..179D}, respectively. Following Sec. 2.1.2 of \citet{2023MNRAS.tmp.3593G}, we remove likely RRL members of LMC and SMC based on their sky positions and proper motions from Gaia DR3 and refined distances from \citet{2023ApJ...944...88L}. We select LMC and SMC stars following three criteria \citep{2023MNRAS.tmp.3593G}. First, we select all stars located within 16$^{\circ}$ and 12$^{\circ}$ from the LMC and SMC, respectively. Second, we require RRLs with proper motions relative to the LMC or SMC less than $\pm$5 mas yr$^{-1}$ based on the proper motion values in \citet{2018A&A...616A...1G}. Third, we select RRLs with $G>18.5$ according to the expected apparent magnitude of the RRLs at LMC and SMC distances \citep{belokurov2017clouds}. For globular clusters, we refer to the catalog of \citet{2010arXiv1012.3224H}, removing the stars within 15 times their half-light radius. 

We use a right-hand Galactocentric Cartesian coordinate system $(X,Y,Z)$, where $X$ points in the direction opposite to the Sun, $Y$ aligns with Galactic rotation within solar neighbourhood and $Z$ points towards the North Galactic Pole. The distance from the Sun to the Galactic center $R_0$ and the local standard of rest (LSR) velocity we adopt are $8.0\,\rm kpc$ \citep{1993ARA&A..31..345R} and $220\,\rm km\,s^{-1}$ \citep{1986MNRAS.221.1023K}, respectively, which match the Galactic potential we choose. The Sun's position is at $(X,Y,Z)=(-R_0, 0, 0)\,\rm kpc$, with a peculiar velocity of $(11.69, 10.16, 7.67)\,\rm km\,s^{-1}$ \citep{2021MNRAS.504..199W}. Tests with alternative solar peculiar velocities \citep[e.g.][]{2010MNRAS.403.1829S,2015MNRAS.449..162H} show minimal impact on our results. We select the Gaia DR3 data with \dt{ruwe}\,$<1.4$ to exclude stars with dubious astrometry \citep{2021A&A...649A...1G}. Based on proper motions, we calculate tangential velocities, $V_l$ and $V_b$, along the Galactic longitude $l$ and latitude $b$ and have converted them to the Galactic standard of rest (GSR) frame based on the solar peculiar velocity and LSR velocity. Stars with each tangential velocity component exceeding $400\,\rm km\,s^{-1}$ or with uncertainties over $100\,\rm km\,s^{-1}$ are excluded, as their velocities might be unreliable. We further utilize the cut $|Z|>3\,\rm kpc$ to to eliminate the majority of the disk and bulge stars. We reserve the stars with $|Z|<3\,\rm kpc$, $R=\sqrt{X^2+Y^2}>20\,\rm kpc$ to identify the parts of substructures in the outer disk.
Finally, our total sample comprises  46,575 halo RRLs (38,825 RRab and 7,750 RRc stars). We plot the distributions of heliocentric distances and metallicities for our sample in Fig.~\ref{fig:allsample_d_FeH}. The majority of our sample covers a heliocentric distance range of 6-49$\,\rm kpc$. The median of the heliocentric distances and metallicities are $20\,\rm kpc$ and $-1.68$, respectively. Fig.~\ref{fig:all_sample_dens} presents the spatial distributions and the distributions of velocities along with the Galactocentric distance $r$, clearly showing the Sgr core and its tidal stream.

\section{Method} %\label{sec:style}
\label{sec:method}

\subsection{Integrals of Motion and Friends-of-friends Algorithm}
\label{subsec:IoM FoF}
Stars within a single substructure are presumed to share similar orbits but possibly at different orbital phases.
In a spherical potential system with no dynamic friction, a star's orbit can be characterized by the IoM. We adopt five IoM parameters: semimajor axis $a$, eccentricity $e$, direction of the orbital pole $(l_{\rm orbit},b_{\rm orbit})$, and the angle between apocenter and the projection of $X$-axis on the orbital plane $l_{\rm apo}$ \citep[][;Xue et al. 2024, in prep.]{2022MNRAS.513.1958W}. These five parameters $(e,a,l_{\rm orbit},b_{\rm orbit}, l_{\rm apo})$ are translated from the total energy $E$ and angular momentum $\boldsymbol{L}$. Notably, $l_{\rm apo}$ is actually not an IoM as it varies with the orbital period but remains constant within one period, making it useful for distinguishing different components within a substructure, such as the leading and trailing arms of the Sgr stream.

Our sample, however, lacks radial velocity measurements, which are crucial for calculating IoM. Therefore, we adopt a prior distribution of radial velocity $V_{\rm los}$ in GSR frame, assuming a Gaussian distribution with the mean at $0\,\rm km\,s^{-1}$ and the velocity dispersion of $109\,\rm km\,s^{-1}$, as estimated from halo RRLs in \citet{2022MNRAS.513.1958W}.
 We then employ the Monte Carlo (MC) method to simulate the distributions of the IoM $\hat{O}=(e,a,l_{\rm orbit},b_{\rm orbit},l_{\rm apo})$. In the MC method, apart from the prior distribution of the radial velocity, the proper motions are randomly sampled by the Gaussian distributions. Since distances cannot be negative, we randomly sample distances based on the logarithmic normal distribution. For the proper motions and distances of stars, it is assumed that the distributions of these quantities follow Gaussian distributions or logarithmic normal distributions with the measured values as means and measurement errors as variances. The right ascension $ra$ and declination $dec$ are very precise, so they are not randomly sampled.
The Galactic potential we adopt is composed of a spherical Hernquist bulge \citep{1990ApJ...356..359H}, an exponential disk and a Navarro-Frenk-White (NFW) halo \citep{1996ApJ...462..563N}. For each star, 10$^5$ sets of $\hat{O}=(e,a,l_{\rm orbit},b_{\rm orbit},l_{\rm apo})$ are calculated and transformed to 5D histogram in IoM space. We uniformly divide the range $0 \leq e \leq 1$ into 10 bins, $0^{\circ} \leq b_{\text{orbit}} \leq 180^{\circ}$ into 10 bins, $0^{\circ} \leq l_{\text{orbit}} \leq 360^{\circ}$ into 20 bins, and $0^{\circ} \leq l_{\text{apo}} \leq 360^{\circ}$ into 20 bins. The bins of $a$ are selected by uniformly dividing $-1 \leq \text{log}_{10}(a/\text{kpc}) \leq 2.4$ into 30 bins. In total, this creates $1.2\times 10^6$ bins. After normalization, we can obtain the orbit-probability-distribution $p(\hat{O})$ for the star. For two stars $i$ and $j$, we define the orbit-likelihood-distance $\mathcal{L}\mathcal{D}_{ij}$ as
\begin{equation}
    \mathcal{L}\mathcal{D}_{ij}=-\ln{\left( 
    \frac{\int{p_i(\hat{O})p_j(\hat{O})d\hat{O}}}{\sqrt{\int{p_i(\hat{O})p_i(\hat{O})d\hat{O}}\times\int{p_j(\hat{O})p_j(\hat{O})d\hat{O}}}}
    \right) }
\end{equation}
to characterize how close two stars locate in IoM space. $p_i(\hat{O})$ and $p_j(\hat{O})$ represent the orbit-probability-distributions of these two stars, respectively.

In the FoF group finding algorithm, two stars will be linked into a group if the dimensionless ``distance'' of these two stars is within a certain threshold, named ``linking length''. In this case, the dimensionless ``distance'' means the orbit-likelihood-distance $\mathcal{L}\mathcal{D}$. This process continues iteratively, adding stars to the group if $\mathcal{L}\mathcal{D}$ between these stars and the two stars meets the criteria, until no new stars can be added, at which point the group construction is complete. 

Tests at different linking lengths show that when the linking length becomes larger, several groups belonging to some more diffuse substructures (like HAC) will merge together, but the relaxed restriction reduces the reliability of the unknown groups. Thus, the numbers of substructure members are observed at varying linking lengths ranging from 0.15 to 0.80 to optimally identify each substructure and discern different substructures with possible common progenitors. 

To determine the critical linking length of each group, we follow the same procedure as Appendix B of \citet{zhang2024VMP}. In most cases, as the linking length increases, the number of members in a group first grows gradually and then exhibits a sudden jump, which may indicate the merging of this group with other group(s). Thus, we take the linking length just before this jump as the critical linking length to select this group, iterating this process accordingly. We manually inspected these jumps and selected Sgr streams first.

To balance the reliability and quantity of identified groups, we select groups with sizes larger than 15. Combined with manual inspection, 220 groups are obtained at appropriate linking lengths.

\begin{deluxetable*}{ccccccccccc}\label{tab:all_sub}
\tabletypesize{\footnotesize}
%\tablenum{3}
\tablecaption{The parameters of the members of all identified known substructures and unknown groups. The errors of the proper motions and the full version of the table are available online.}
\tablewidth{0pt}
\tablehead{
source\_id (Gaia DR3) & ra & dec  & $d$ & $d_{\rm err}$ & $\rm [Fe/H]$ & $\rm [Fe/H]_{err}$ & pmra  & pmdec & vartype  & label\\
 & $\rm (deg)$ & $\rm (deg)$ & $\rm (kpc)$ & $\rm (kpc)$ &  & $\rm (dex)$ & $(\rm mas\,yr^{-1})$  & $(\rm mas\,yr^{-1})$ &  &   }
\startdata
4128577965154004352 & 257.6824 & -18.8837 & 32.58 & 3.05 & -1.91 & 0.31 & -1.463 & -1.315 & RRab & Sgr leading\\ 
4130100647016394496 & 249.5106 & -21.1458 & 32.63 & 3.71 & -2.15 & 0.45 & -1.478 & -1.145 & RRab & Sgr leading\\ 
4132532320062978176 & 253.1924 & -17.6400 & 27.77 & 2.85 & -1.17 & 0.38 & -1.482 & -0.902 & RRab & Sgr leading\\ 
4131954599718690688 & 250.6789 & -17.6604 & 27.78 & 3.20 & -1.22 & 0.51 & -0.932 & -0.930 & RRab & Sgr leading\\ 
4132128936735267584 & 252.6425 & -18.9909 & 29.45 & 3.15 & -1.69 & 0.42 & -1.416 & -0.955 & RRab & Sgr leading\\ 
4323521166298784256 & 247.9425 & -18.6516 & 28.05 & 2.87 & -1.56 & 0.39 & -1.677 & -0.791 & RRab & Sgr leading\\ 
4131289773140347904 & 248.1501 & -19.5035 & 25.29 & 2.49 & -1.92 & 0.32 & -1.805 & -0.826 & RRab & Sgr leading\\ 
4130828974686677504 & 252.0072 & -18.9382 & 32.16 & 3.09 & -1.71 & 0.32 & -1.070 & -1.053 & RRab & Sgr leading\\ 
4130970738671045376 & 249.4351 & -20.2542 & 36.17 & 3.71 & -1.74 & 0.37 & -0.955 & -0.852 & RRab & Sgr leading\\ 
4130362330788414848 & 250.3936 & -19.5540 & 24.97 & 2.94 & -1.21 & 0.51 & -1.998 & -0.610 & RRab & Sgr leading\\  
\enddata
\end{deluxetable*}

\subsection{Methodology Validation}

\begin{figure}
	\centering
	\includegraphics[width = \linewidth]{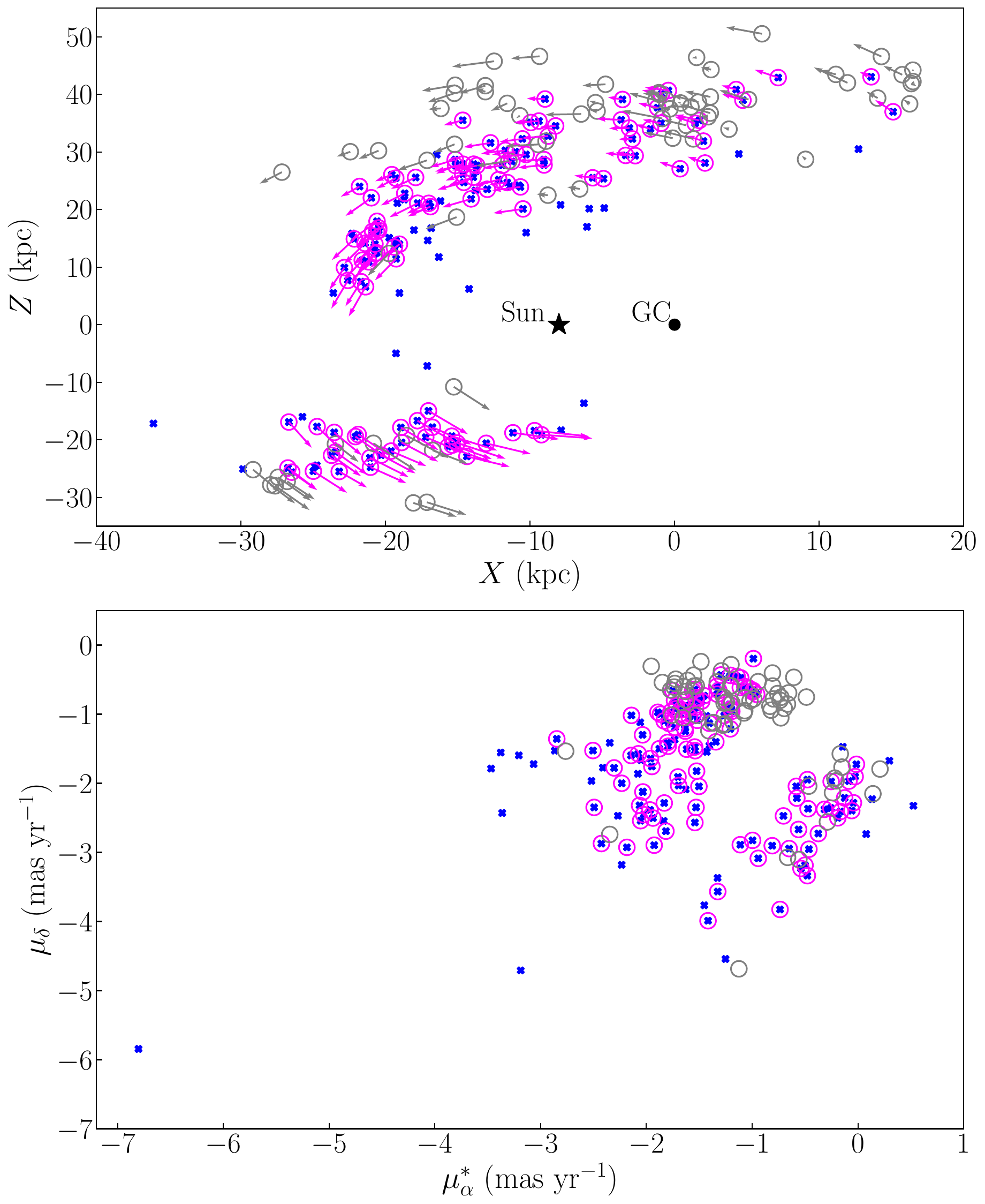}
	\caption{Comparisons with the Sgr members of \citet{2022MNRAS.513.1958W} in the $(X,Z)$ and $(\mu_\alpha^*,\mu_\delta)$ space, respectively. The blue symbol X marks the Sgr members identified by \citet{2022MNRAS.513.1958W} using 6D information. The open circles represents the results of our method by utilising 5D information and the colors magenta and gray mean stars in common with the result of \citet{2022MNRAS.513.1958W} and possible contaminations, respectively. The arrows in the top panel corresponding with the open circles show the moving directions and velocity amplitudes of these RRLs.}
	\centering
	\label{fig:wang2022}
\end{figure}

The members of a stellar stream typically share similar orbital properties, making them relatively easier to identify. So we validate our method using the known Sgr stream. \citet{2022MNRAS.513.1958W} identified 145 RRLs belonging to the Sgr streams from $\sim$3,000 RRab stars with full velocity information. Using the same dataset except for the radial velocities, we identify the Sgr stream and compare the two results to validate our method. We only keep groups with at least 5 members. The critical linking length is determined as described in Sec.~\ref{subsec:IoM FoF}. We recognize 177 RRLs likely belonging to the Sgr stream, of which 110 RRLs are in common with the result of \citet{2022MNRAS.513.1958W}. It shows we still can identify about 76\% of Sgr stream members based on 5D kinematic parameters, comparing to the case of having full-phase space information. Fig.~\ref{fig:wang2022} illustrates the comparisons and shows that most misidentifications are similar to other Sgr stars in the spatial distribution and proper motion space. The velocity vector arrows in the top panel of Fig.~\ref{fig:wang2022} present the different velocities between common ones (magenta open circles) and possible contamination ones (grey open circles), suggesting that these misidentifications likely stem from the absence of radial velocity information. 
According to our results, the lack of radial velocity leads to a misidentification of about 38\%.

We also utilize the Sgr stream in our sample to validate our method by comparing the large RRL catalogue of Sgr candidates from \citet{2020A&A...638A.104R}. After cross-matching their catalog with our total sample, we find 5,166 Sgr candidates. We recognize 4,898 RRLs belonging to the Sgr of which 4,215 RRLs are in common with the Sgr candidates of \citet{2020A&A...638A.104R}, which implies that the completeness and purity of our identified Sgr RRL members are approximately 82\% and 86\%, respectively. Actually, the values of completeness and purity are slightly overestimated because we assume their sample is complete and pure.
This completeness is consistent with the validation by using $\sim$3,000 RRab with full velocity information. The higher purity is caused by the Sgr candidates we compare which are selected based on the positions and proper motions. These validations indicate that our method can accurately distinguish about four-fifths of a substructure's members, with most misidentifications attributable to the lack of radial velocities.

\begin{figure*}
	\centering
	\includegraphics[width = \linewidth]{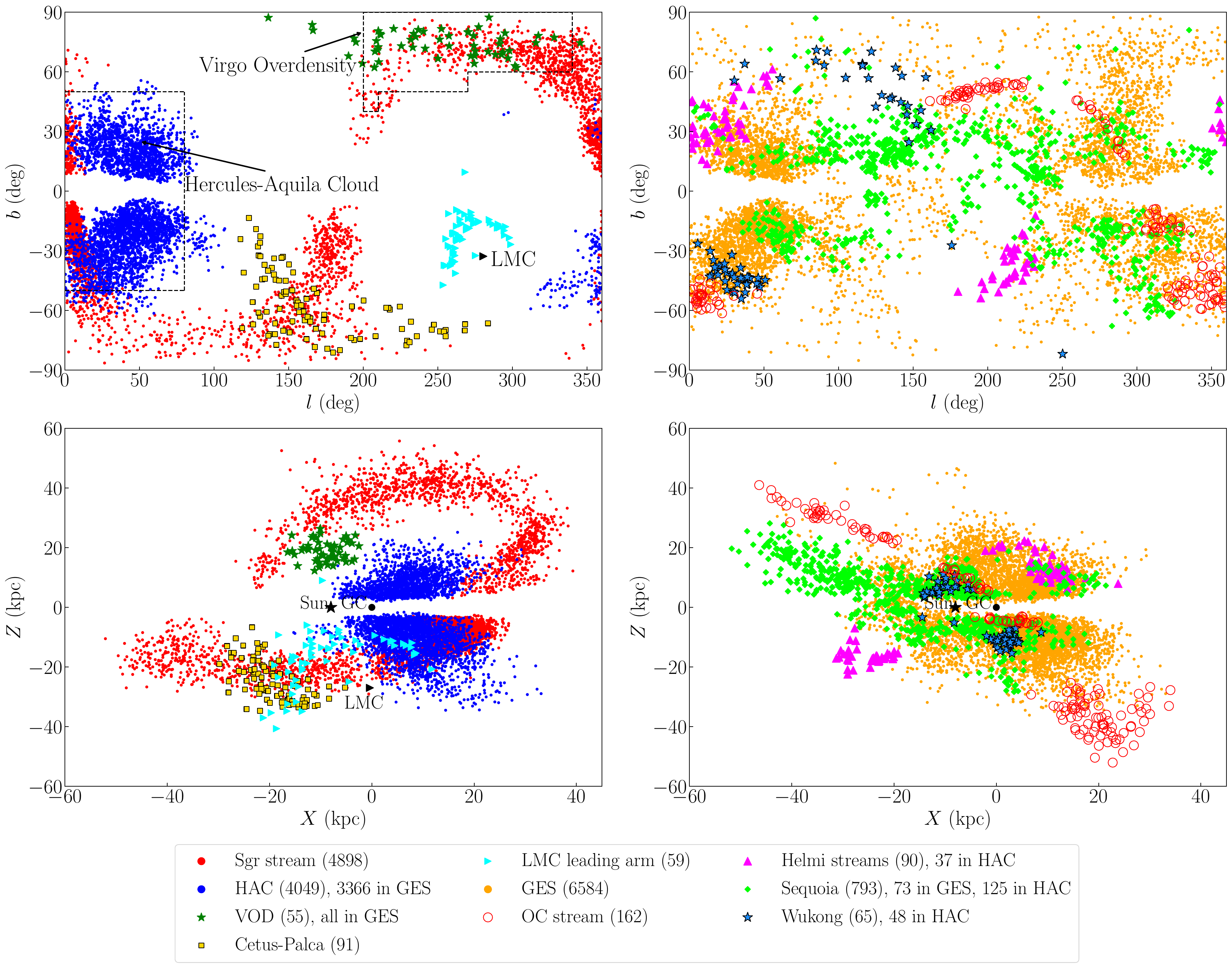}
	\caption{The spatial distributions of the all known substructures identified by our RRL sample in the $(l,b)$ and $(X,Z)$ spaces. These substructures are shown in two columns of panels for clarity. The numbers in parentheses are the number of candidates for each substructure. The black dashed areas in the upper left panel represent the regions of HAC and VOD in Table~\ref{tab:diffuse_sub}. The black stars and black dots in the bottom two panels present the position of the Sun and the Galactic center, respectively. The black triangle in the left two panels represent the position of LMC center. The overlap between several pairs of substructures in Table~\ref{tab:overlap} is listed in the legend and can be directly observed.}
	\centering
	\label{fig:all_known_lbXZ}
\end{figure*}

\section{Results}\label{sec:results}

As described in Sec.~\ref{sec:method}, we identify substructures with the FoF method in the IoM space by adopting a prior distribution of the radial velocity. We identify several known substructures: the Sgr stream and Sgr core \citep{2001ApJ...551..294I,2003ApJ...599.1082M}, the Hercules–Aquila Cloud \citep[HAC;][]{2007ApJ...657L..89B}, the Virgo Overdensity \citep[VOD;][]{2002ApJ...569..245N}, the Gaia-Enceladus-Sausage \citep[GES;][]{2018MNRAS.478..611B,2018Natur.563...85H}, the Helmi Streams \citep{1999Natur.402...53H,2019A&A...625A...5K,2019A&A...631L...9K}, Sequoia \citep{2018MNRAS.478.5449M,2019MNRAS.488.1235M,2019ApJ...874L..35M}, Wukong \citep{2020ApJ...901...48N}, Cetus-Palca \citep{2020ApJ...905..100C,2022A&A...660A..29T} and the Orphan-Chenab (OC) stream \citep{2006ApJ...645L..37G,2007ApJ...658..337B,2019MNRAS.485.4726K,2019ApJ...885....3S}. A group is identified to be associated with LMC leading arm. These known substructures we identify could also demonstrate the reliability of our approach.
Fig.~\ref{fig:all_known_lbXZ} shows the spatial distributions of all substructures identified in our RRL sample. Apart from known substructures, 18 remaining groups are identified to be not associated with any known substructures. Table~\ref{tab:all_sub} lists the parameters of the members of all identified substructures and unknown groups.

Utilizing the same RRL catalog of \citet{2023ApJ...944...88L} as this work, \citet{2023MNRAS.tmp.3593G} identified several recognized substructures of the MW including Sgr stream, GES, Helmi streams, Sequoia and Wukong. Our results cover these substructures and  present an order of magnitude more candidates than \citet{2023MNRAS.tmp.3593G}.

\begin{figure*}
	\centering
	\includegraphics[width = \linewidth]{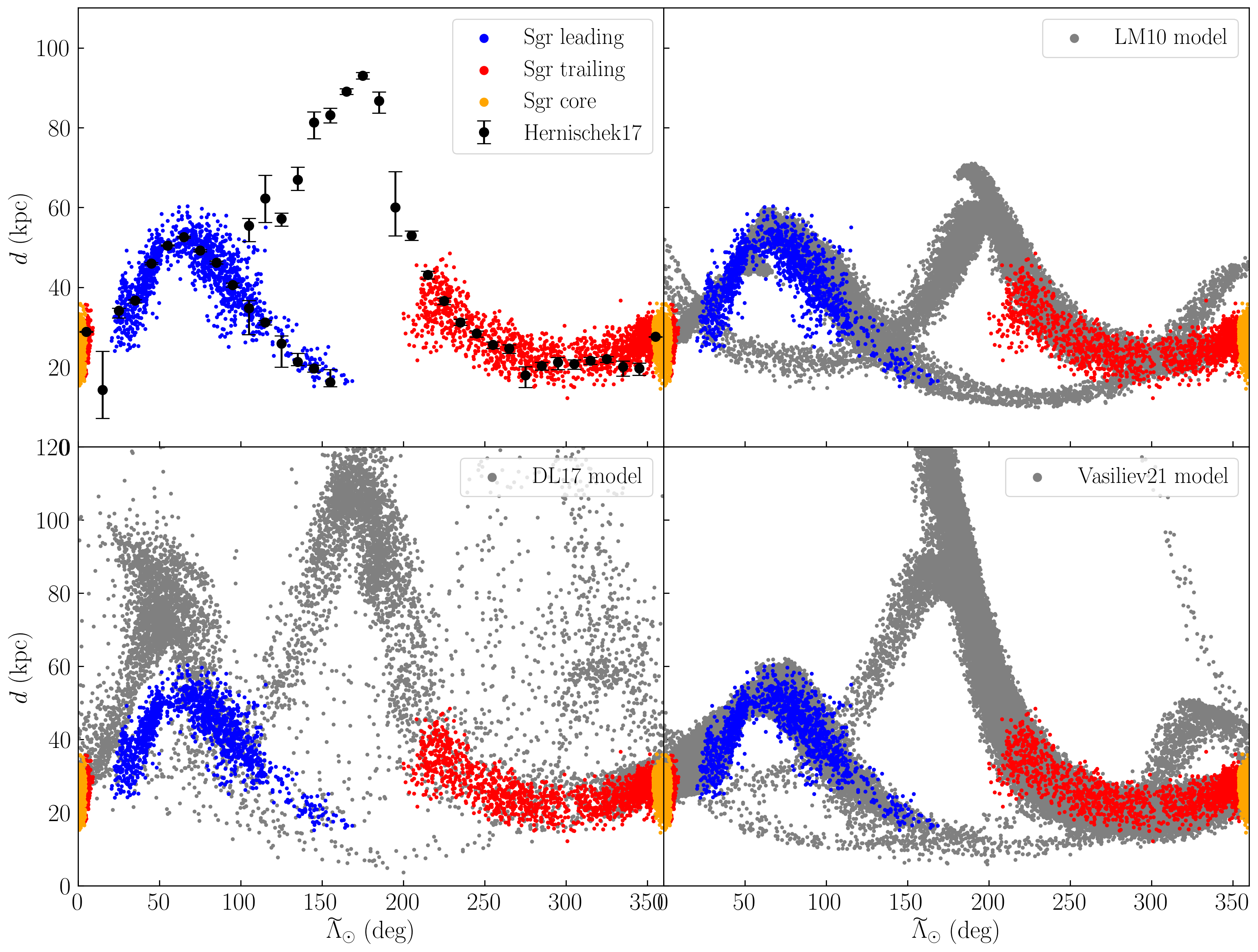}
	\caption{Comparisons with the observation result, LM10 model and DL17 model in coordinates of $(\widetilde{\Lambda}_\odot,d)$. The blue, red, and orange dots represent members of Sgr leading arm, Sgr trailing arm, and Sgr core, respectively. The black dots with error bars in the top panel are from Tables 4 and 5 of \citet{2017ApJ...850...96H} obtained from RR Lyrae stars. The gray dots in the upper right and bottom panels indicate the LM10 model, DL17 model, and Vasiliev21 model, respectively.}
	\centering
	\label{fig:Sgr_compare_models}
\end{figure*}

\begin{figure*}
	\centering
	\includegraphics[width = \linewidth]{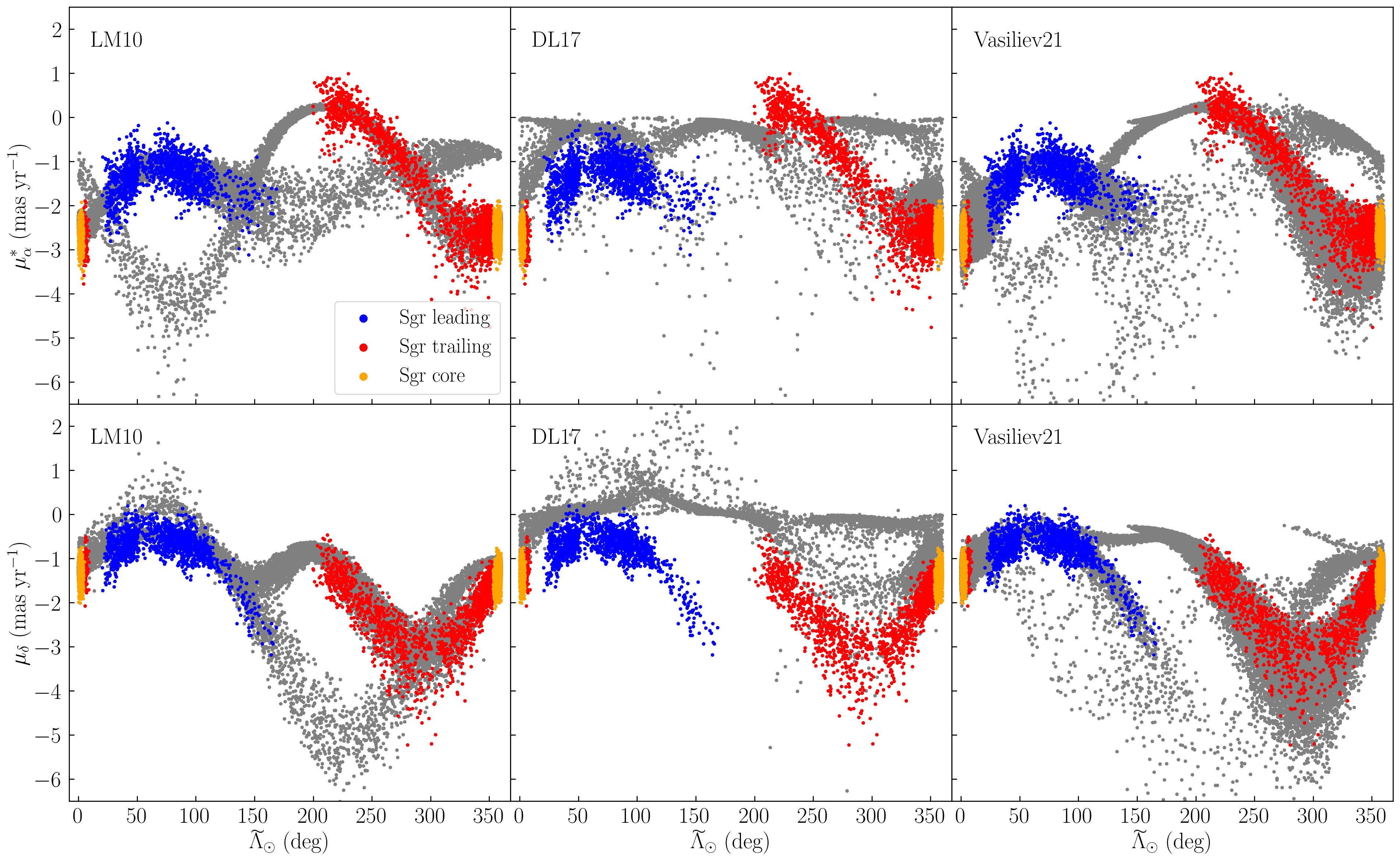}
	\caption{The proper motion distributions of our Sgr RRLs along with $\widetilde{\Lambda}_\odot$. The blue, red, and orange dots represent the RRLs belonging to the Sgr leading arm, Sgr trailing arm, and Sgr core, respectively. The gray dots in the left, middle and right panels are from the DL17 model, LM10 model, and Vasiliev21 model, respectively.}
	\centering
	\label{fig:Sgr_compare_models_pm}
\end{figure*}

\begin{figure}
	\centering
	\includegraphics[width = \linewidth]{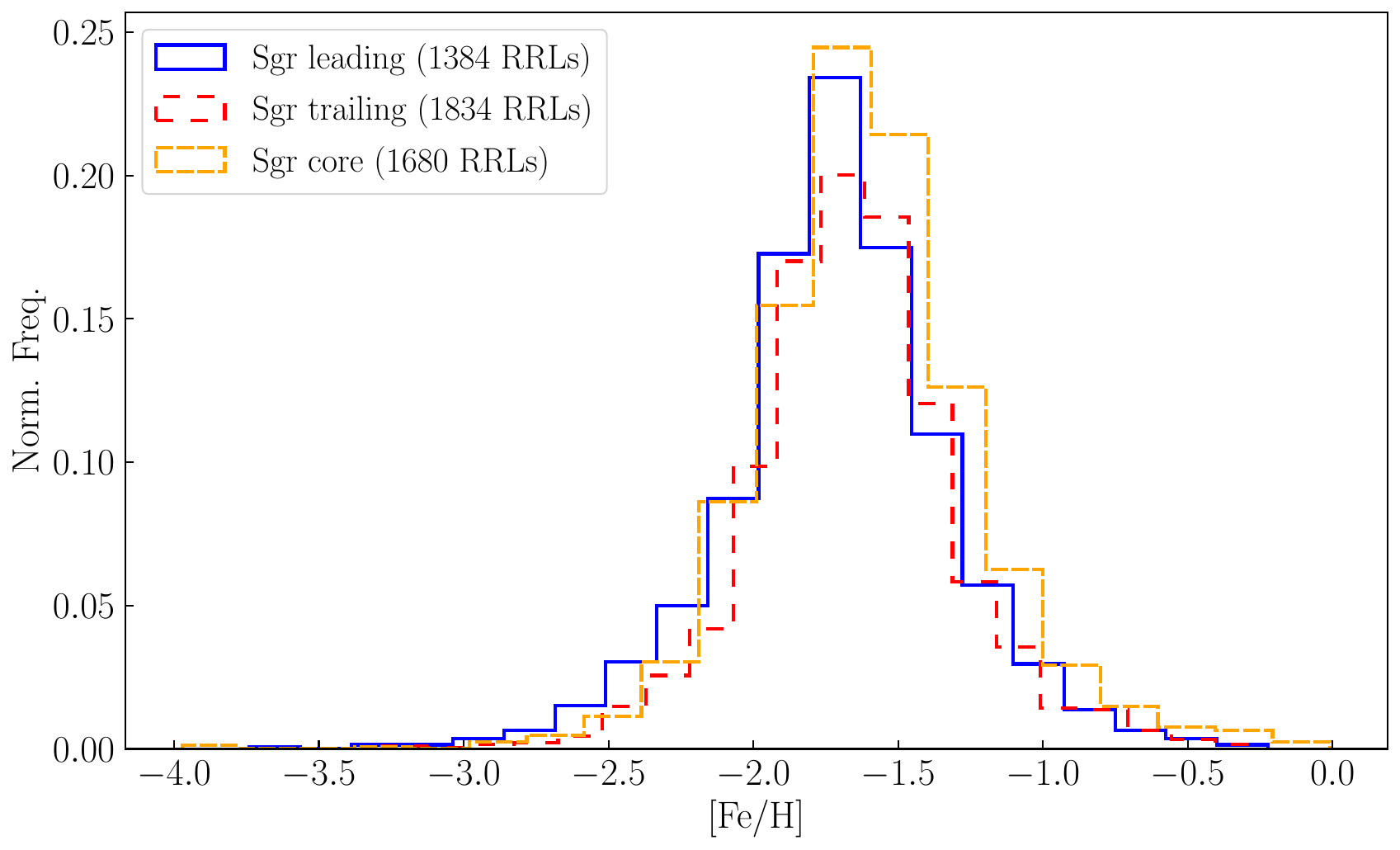}
	\caption{The metallicity distributions of the RRLs belonging to the Sgr leading arm, Sgr trailing arm, and Sgr core, respectively.}
	\centering
	\label{fig:Sgr_metallicity}
\end{figure}

\subsection{Known substructures}
The spatial distribution of stellar streams is concentrated, and their tangential velocities are well-clustered. Also, the identification of overdensities with well-defined spatial distributions is less affected by the lack of radial velocities. So these two types of substructures can be reliably identified by comparison with samples from the literature or the \texttt{galstreams} library \citep{2023MNRAS.520.5225M} in both three-dimensional positions and two-dimensional proper motion space. 

In contrast, other substructures that are less clustered in spatial and proper motion distributions but better clustered in energy, angular momentum, or action space may be more significantly impacted by the absence of radial velocities. Additionally, due to their diffuse spatial distribution, it is challenging to directly validate the identification efficiency through cross-matching with literature.

To apply a uniform method of quantitatively estimating the probability of our RRL groups being associated with known substructures, we perform 100 Monte Carlo simulations for each star with radial velocity prior in a group, following the same procedure as in Sec.~\ref{subsec:IoM FoF}. Then all the MC results are concatenated for all stars in a group. The probability or ``confidence'' of a group being associated with a particular substructure is calculated by dividing the number of MC result points that fall within the criteria of selecting the substructure by the total number of MC result points of the group. 

For streams including Sgr stream, OC stream and Cetus-Palca, groups with high association confidence can be directly selected by eye. For overdensities and spatially diffused substructures, an appropriate confidence threshold is determined through individual inspection and verification with member stars that have Gaia or LAMOST radial velocities. Only the groups with an association confidence higher than this threshold are selected as candidate groups of that substructure. 
So our procedure can give candidate members of all three common types of known substructures: streams, overdensities and spatially diffused substructures.
When comparing with literature criteria, it is worth noting that the Galactic potential we adopt in Sec~\ref{subsec:IoM FoF} results in our mean orbital energy $E$ being $0.05\times10^5$ km$^2$s$^{-2}$ higher than that of orbits under the \texttt{MWPotential2014} \citep{bovy2015galpy} by test, which was adopted by \citet{2020ApJ...901...48N}.

\subsubsection{Sgr Stream: model validation and metallicity gradients}\label{Sec:Sgr}
Sgr stream is the most prominent substructure in our Galactic halo with distinct spatial distribution (see Fig.~\ref{fig:all_sample_dens}). We compare our groups with the fitting results of Sgr stream in the coordinate of $(\widetilde{\Lambda}_\odot,d)$ from \citet{2017ApJ...850...96H}. $\widetilde{\Lambda}_\odot$ is the longitude in the Sgr coordinate system and the definition is the same as that in \citet{2014MNRAS.437..116B}. We find three groups (4,898 RRLs) are consistent with the Sgr streams. In the three groups, two (1,384 RRLs) belong to the Sgr leading arm and another (3,514 RRLs) belongs to the Sgr trailing arm and the Sgr core. We extract the stars within the Sgr core region of $279^{\circ}<\alpha<291^{\circ}$ and $-40^{\circ}<\delta<-20^{\circ}$ \citep{2020MNRAS.495.4124F}. 
As shown in Fig.~\ref{fig:Sgr_compare_models}, the distributions of our Sgr RRLs in the $(\widetilde{\Lambda}_\odot,d)$ space are highly consistent with the fitting results of \citet{2017ApJ...850...96H}. In addition, the Sgr stream is well clustered into three groups, which suggests that the two components of Sgr stream can be well recognized by our method even without the radial velocity information.

In assessing various Sgr models, we compare our large sample of Sgr members to the models proposed by \citet{2010ApJ...714..229L} (hereafter LM10), \citet{2017ApJ...836...92D} (hereafter DL17), and \citet{2021MNRAS.501.2279V} (hereafter Vasiliev21). As illustrated in Figs.~\ref{fig:Sgr_compare_models} and~\ref{fig:Sgr_compare_models_pm}, the distance and the proper motion of the LM10 and Vasiliev21 models are more consistent with our Sgr RRLs, suggesting that both models could reconstruct the leading and trailing arms. Besides, these two models show different features in the $100^\circ<\widetilde{\Lambda}_\odot<200^\circ$. \citet{2019ApJ...886..154Y} found that in the range of $100^\circ<\widetilde{\Lambda}_\odot<200^\circ$ and $d>60\,\rm kpc$, the radial velocity and distance of LM10 model cannot match with the Sgr debris component traced by the K giants, M giants and BHB stars, indicating that the Vasiliev21 model is more consistent with observations. Our study does not include Sgr debris components in this region due to the removal of stars with large velocity uncertainties, so we cannot validate this part of the LM10 and Vasiliev21 models in our work. Moreover, our Sgr RRLs are more consistent with LM10 and Vasiliev21 models in $(\widetilde{\Lambda}_\odot,d)$ space than \citet{2017ApJ...850...96H} which exhibits a dip around $\widetilde{\Lambda}_\odot\sim330^{\circ}$ (see Fig.~\ref{fig:Sgr_compare_models}), serving as additional validation of our identification of the Sgr stream.

\begin{figure*}
	\centering
	\includegraphics[width = \linewidth]{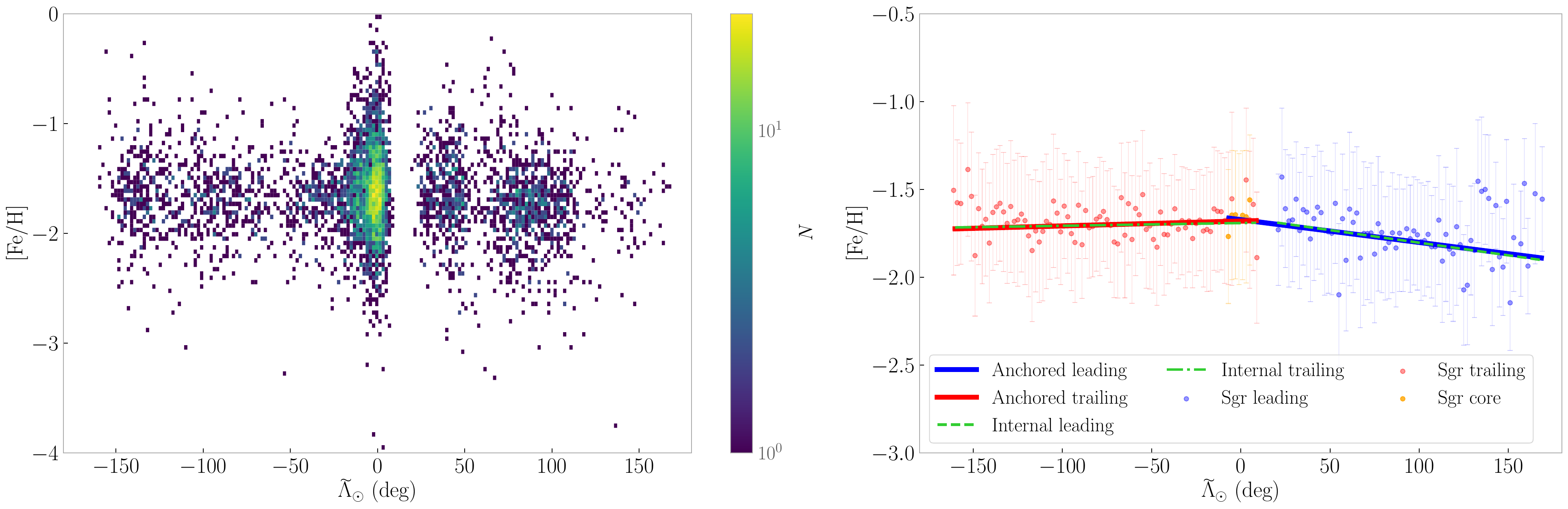}
	\caption{Left: Metallicity versus Sgr longitude $\widetilde{\Lambda}_\odot$ in a 2D histogram. Right: The metallicities of Sgr leading arm (blue points), trailing arm (red points) and core (orange points) binned in $\widetilde{\Lambda}_\odot$, with errorbars indicating the mean uncertainty in each bin. Assuming linear metallicity trends along each arm, the solid and dashed lines representing the ``anchored'' and ''internal'' fittings, respectively. In both panels, $\widetilde{\Lambda}_\odot$ is transformed from $0^{\circ}\sim 360^{\circ}$ to $-180^{\circ}\sim 180^{\circ}$ for a continuous display of both arms.}
	\centering
	\label{fig:Fig_Sgr_metal}
\end{figure*}

We compare the metallicities of three Sgr components using our RRL sample. The metallicity distributions of these three Sgr components are shown in Fig.~\ref{fig:Sgr_metallicity}. The mean metallicities of the leading arm, trailing arm, and core are $-1.71$, $-1.65$, and $-1.60$, respectively, and the corresponding standard deviations are 0.39, 0.35 and $0.40\,\rm dex$, respectively. The metallicity results align with the work of \citet{2022MNRAS.513.1958W} who identified the metallicities of $-1.70$ and $-1.64$ in the leading arm and trailing arm, respectively,  using RRLs with spectroscopic metallicity measurements. The Sgr metallicities are also consistent with Table 5 of \citet{2023MNRAS.tmp.3593G}, which utilizes the same RRL catalog of \citet{2023ApJ...944...88L}. Using metal-rich M giants, \citet{2018ApJ...859L..10C} found that the average metallicity of the Sgr stream was lower than the Sgr core. Our results of RRLs corroborate this (see Fig.~\ref{fig:Fig_Sgr_metal}), showing the core to be relatively metal-rich compared to the arms, though the differences are not significant. 

Fig.~\ref{fig:Fig_Sgr_metal} plots the distribution and gradients along the whole Sgr stream we identify. Assuming linear metallicity treads along Sgr leading arm and trailing arm respectively, we fit the ``internal'' (only including the corresponding Sgr stream) and ``anchored'' (including the particular Sgr stream and core members) cases following \citet{hayes2020Sgr}. The gradients are listed in Table~\ref{tab:Sgr_metal}, which are the first measurements of Sgr stream metallicity gradients by RRLs.  In the trailing arm, our result is consistent (within 1$\sigma$) with no internal metallicity gradient. Considering the re-estimated photometric metallicity uncertainties of $0.24\,\rm dex$ for RRab and $0.16\,\rm dex$ for RRc, respectively, in \citet{2023ApJ...944...88L}, the metallicity gradient detected in Sgr leading arm is significant.

\begin{deluxetable}{cc}\label{tab:Sgr_metal}
\tabletypesize{\footnotesize}
%\tablenum{3}
\tablecaption{Metallicity gradients along the Sgr stream.}
\tablewidth{0pt}
\tablehead{
& Metallicity gradient (deg$^{-1}$) }
\startdata
Internal leading &   $(1.4\pm0.3)\times 10^{-3}$\\
Internal trailing &  $(0.2\pm0.2)\times 10^{-3}$\\
Anchored leading &   $(1.3\pm0.2)\times 10^{-3}$\\
Anchored trailing &  $(0.3\pm0.1)\times 10^{-3}$\\
\enddata
\end{deluxetable}

\begin{figure*}
	\centering
	\includegraphics[width = \linewidth]{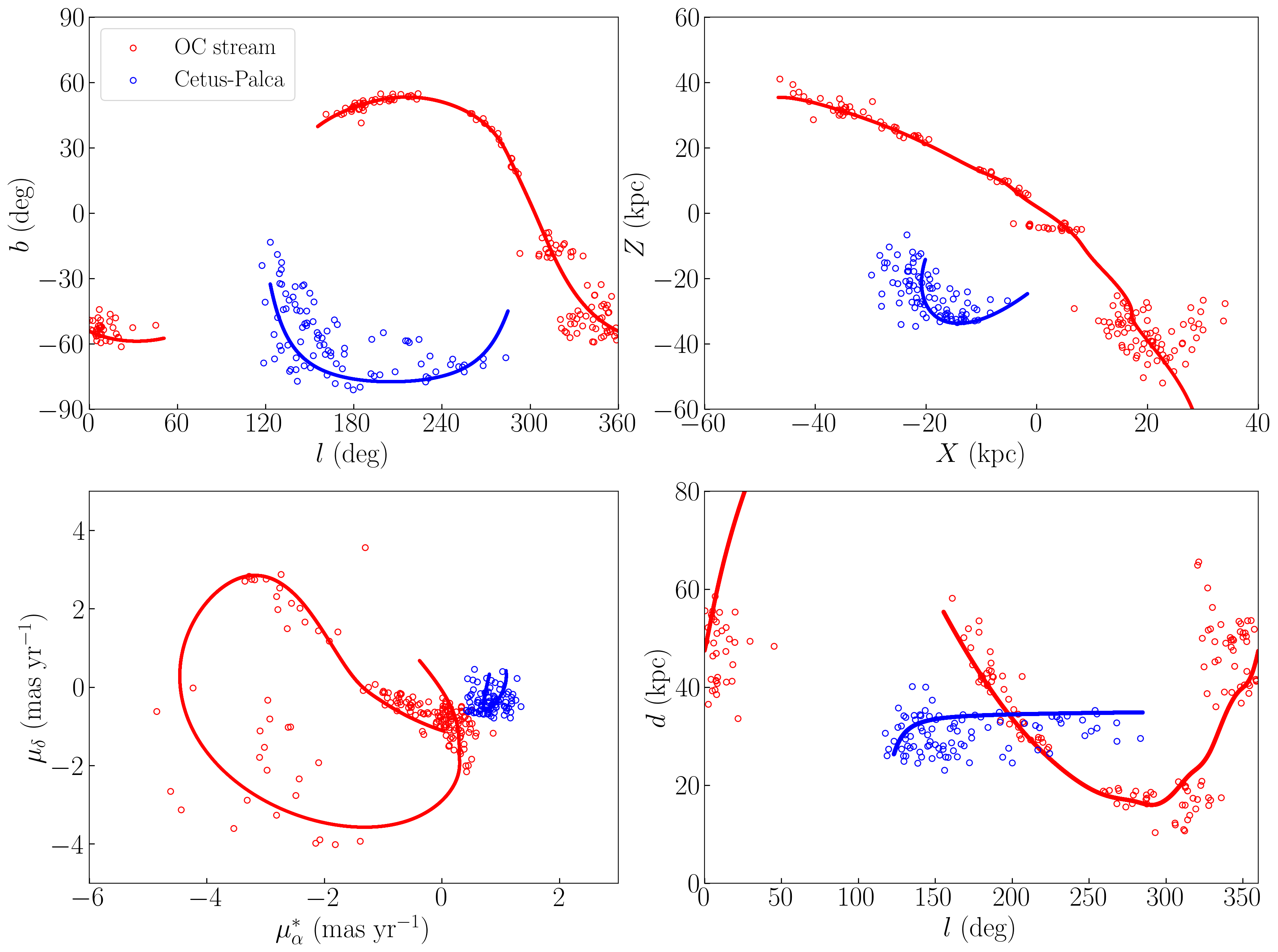}
	\caption{The spatial distribution and proper motion of OC stream (red) and Cetus-Palca (blue). The open circles are the groups identified to be associated with these two substructures, respectively. The solid lines represent the track of Orphan-Chenab and Cetus-Palca in the \texttt{galstreams} library \citep{2023MNRAS.520.5225M}.}
	\centering
	\label{fig:OC_Cetus}
\end{figure*}

\subsubsection{Attributing Groups to Orphan-Chenab Stream and Cetus-Palca}
\begin{figure}
	\centering
	\includegraphics[width = \linewidth]{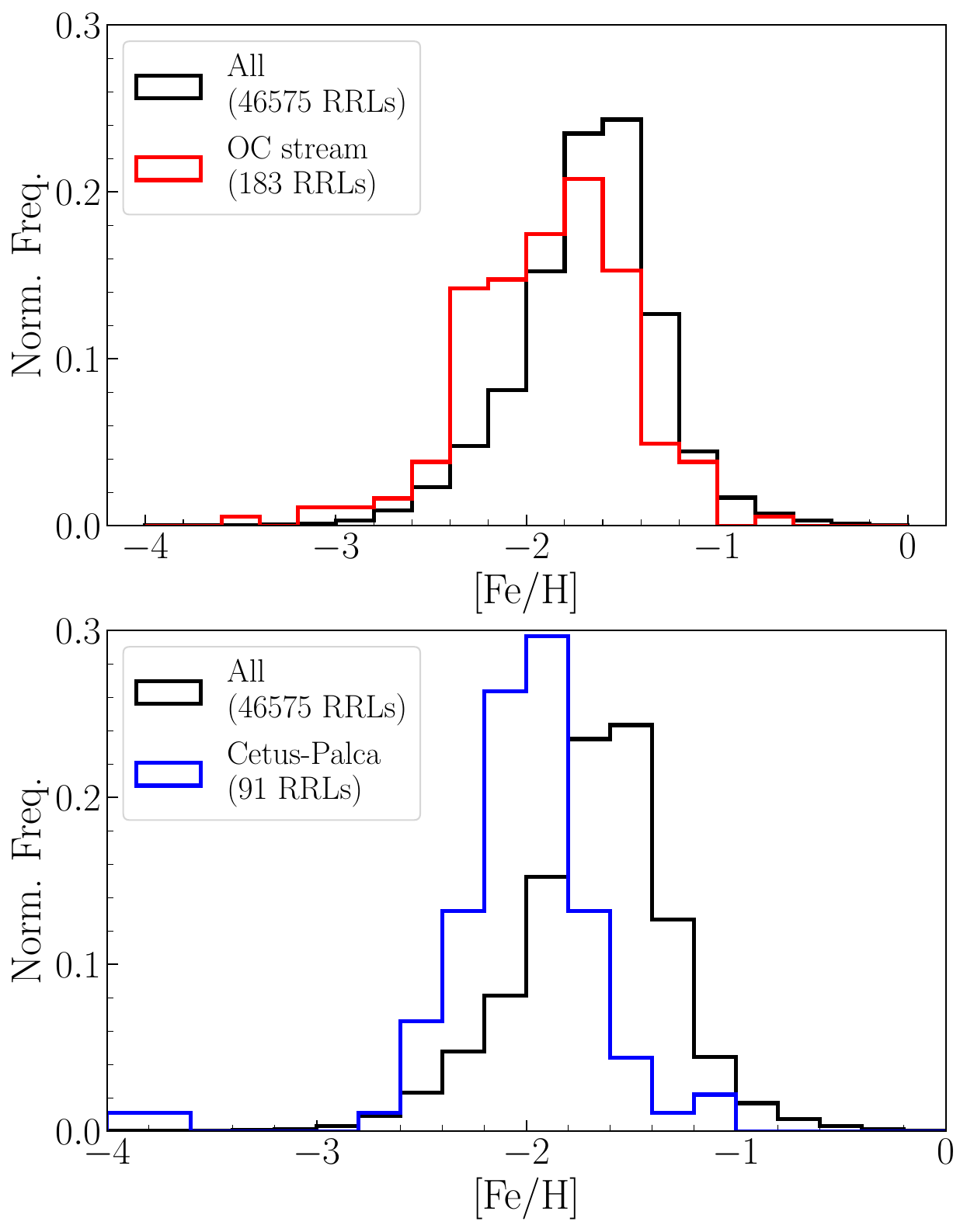}
	\caption{The metallicity distributions of the OC stream (top panel) and Cetus-Palca (bottom panel), shown with red and blue lines, respectively. The black lines in both panels represent the total sample. }
	\centering
	\label{fig:OC_Cetus_FeH}
\end{figure}

Apart from the Sgr stream, we also compare our groups with other known stellar streams in the \texttt{galstreams} library. 
We find 4 groups (162 RRLs) associated with the OC stream and one group (91 RRLs) corresponding to Cetus-Palca, which are consistent with the tracks in \texttt{galstreams} library (see Fig~\ref{fig:OC_Cetus}) in 3D position and proper motion spaces. The tracks of the OC stream and Cetus-Palca were respectively fitted by \texttt{galstreams} from the samples of \citet{2019MNRAS.485.4726K} and \citet{2022A&A...660A..29T}. The portion of the OC stream south of the Galactic plane is relatively diffuse, which may be due to the gravitational perturbation from the LMC \citep{LMC_OC2019}. Fig~\ref{fig:OC_Cetus_FeH} gives the metallicity distributions of OC stream and Cetus-Palca RRLs. The mean and standard deviation of metallicities are $-1.88$ and $0.42\,\rm dex$ for RRLs belonging to the OC stream, consistent with the metallicities of $-1.85\pm0.53\,\rm dex$ for Orphan stream and $-1.78\pm0.34\,\rm dex$ for Chenab stream in \citet{2022ApJ...926..107M}.

The streams we identify reach as far as 50-60 kpc due to the broad distance distribution of our RRL sample. The identification of these stellar streams further demonstrates the effectiveness of our method for detecting streams in the IoM space in the absence of radial velocities. The reason for not identifying more known stellar streams may be the lack of radial velocities, the insufficient proper motion information provided by the \texttt{galstreams} library, or the choice of parameters in our IoM space.

\begin{deluxetable*}{ccc}\label{tab:diffuse_sub}
\tabletypesize{\footnotesize}
%\tablenum{3}
\tablecaption{Selection criteria for some spatially diffused substructures.}
\tablewidth{0pt}
\tablehead{
Substructures & Selection criteria \\  }
\startdata
HAC &  $ 0^{\circ }< l<80^{\circ };\; -50^{\circ }< b< 50^{\circ }$   \\
VOD &   $200^{\circ }< l<210^{\circ }, \; b>40^{\circ }$ or $210^{\circ }< l<270^{\circ }, \; b>50^{\circ }$ or $270^{\circ }< l<340^{\circ }, \; b>60^{\circ }$    \\
GES & $V_\perp<60\,\rm km\,s^{-1};\;r>15\,\rm kpc$     \\
Helmi streams & $1.6<L_{\perp}<3.2\,(\times10^3\;\mathrm{kpc\;km\;s^{-1}}) ;\; -1.7<L_z<-0.75\,(\times10^3\;\mathrm{kpc\;km\;s^{-1}})$  \\ 
Sequoia & $\eta>0.15 ;\; L_z>0.7\,(\times10^3\;\mathrm{kpc\;km\;s^{-1}}) ;\; E>-1.2\,(\times 10^5 \mathrm{~km}^2 \mathrm{~s}^{-2})$ \\
Wukong & $-1<L_z<-0.2\,(\times10^3\;\mathrm{kpc\;km\;s^{-1}}) ;\; -1.30<E<-0.85 \,(\times10^5\;\mathrm{km^2\;s^{-2}});\; \left(J_z-J_R\right) / J_{\text {tot }}>0.3 ;\; 90^{\circ}<\theta<120^{\circ}$
\enddata
\tablecomments{References of the selection criteria: HAC \citep{2007ApJ...657L..89B}; VOD \citep{2012AJ....143..105B}; Helmi Stream \citep{2020ApJ...901...48N}; Sequoia/Arjuna/I'itoi \citep{2020ApJ...901...48N}; Wukong \citep{2020ApJ...901...48N,2024MNRAS.530.2512L}. 
}
\end{deluxetable*}

\subsubsection{Overdensities: Hercules–Aquila Cloud and Virgo Overdensity}\label{Sec:HAC_VOD}

The HAC is an overdensity found by \citet{2007ApJ...657L..89B} using main-sequence and turnoff stars selected from the SDSS DR5. The authors suggested that it was centered on the Galactic longitude $l{\sim}40^{\circ}$ and Galactic latitude $-50^{\circ}<b<50^{\circ}$. Several works have used RRLs to study the properties of HAC \citep{2009MNRAS.398.1757W,2010ApJ...708..717S}. \citet{2014MNRAS.440..161S} mapped the HAC using $\sim14,000$ RRLs from the Catalina Sky Survey \citep{2014ApJS..213....9D}, noting a strong excess in the region of the sky approximately in $30^{\circ}<l<55^{\circ}$ and $-45^{\circ}<b<-25^{\circ}$, and located at a distance of 10-25 kpc away from the Sun. Besides, \citet{2019ApJ...880...65Y} found that the HAC members were located in a heliocentric distance range of 12-21$\,\rm kpc$ and the mean radial velocity was $33.37\,\rm km\,s^{-1}$ using K giants from LAMOST DR5. 

\begin{figure*}
	\centering
	\includegraphics[width = \linewidth]{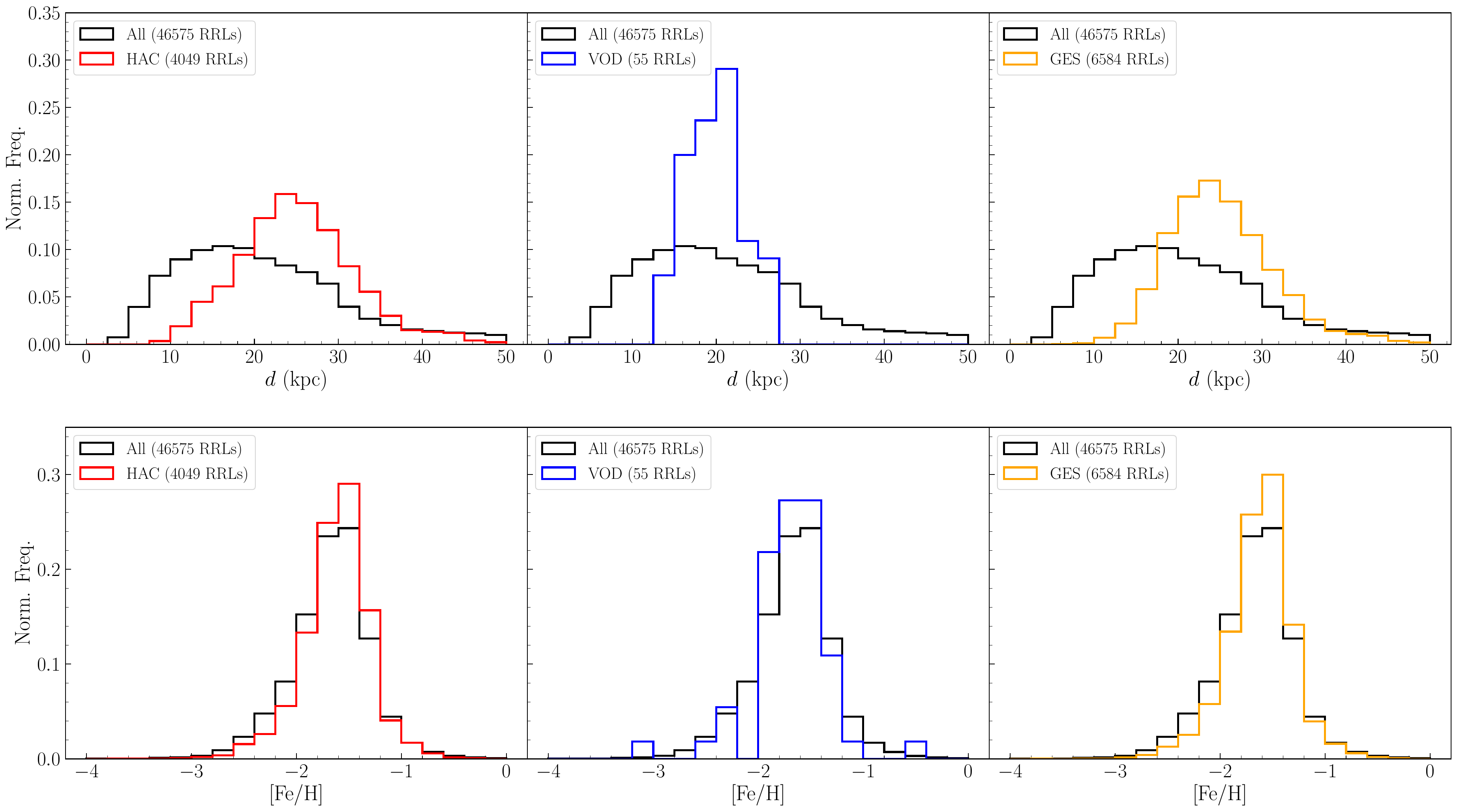}
	\caption{The heliocentric distance and metallicity distributions of the total sample and the members of HAC, VOD, and GES.}
	\centering
	\label{fig:d_FeH_HAC_VOD_GES}
\end{figure*}

According to Table~\ref{tab:diffuse_sub}, 42 groups (4,049 RRLs) are identified as the HAC candidates. The division of the HAC into several groups in our results is attributed to the lack of radial velocity information. Fig.~\ref{fig:d_FeH_HAC_VOD_GES} illustrates the heliocentric distance and metallicity distributions of HAC members. We find the heliocentric distance distribution of HAC members is different from our total sample, highlighting a significant overdensity.
The range and peak of the heliocentric distances are 13-42$\,\rm kpc$ and $\sim 25\,\rm kpc$, respectively, estimated by the 2.5\%, 97.5\%, and 50\% quantile. The metallicity analysis reveals no marked difference between the HAC members and the total sample, with mean and standard deviation values of $-1.62$ and $0.33\,\rm dex$, respectively. Similar to \citet{2014MNRAS.440..161S}, we also find that the spatial distributions of the HAC members are not symmetric, with the majority (2,499 RRLs) in the Galactic Southern hemisphere (see Fig.~\ref{fig:all_known_lbXZ}).

The Virgo Overdensity is a high stellar density region identified by \citet{2002ApJ...569..245N} using turnoff stars from SDSS. The VOD spans over $1,000\,\rm deg^2$ and is located at a heliocentric distance of 10 to 20 kpc \citep{2007ApJ...668..221N,2008ApJ...673..864J,2012AJ....143..105B,2014A&A...566A.118D}. But the origin of this overdensity is still ambiguous. 

There are 2 groups (55 RRLs) consistent with VOD satisfying the criteria in Table~\ref{tab:diffuse_sub}.
As depicted in Fig.~\ref{fig:d_FeH_HAC_VOD_GES}, this overdensity is also significant. It lies 14-26$\,\rm kpc$ away from the Sun and the peak of the heliocentric distance is $20\,\rm kpc$, which are estimated by the 2.5\%, 97.5\%, and 50\% quantile. The mean and standard deviation of the metallicities of these groups are $-1.68$ and $0.37\,\rm dex$, respectively. 

\subsubsection{Attributing Groups to Gaia-Enceladus-Sausage}
\label{subsec:GES}
Using a large sample of main-sequence stars within ${\sim}10\,\rm kpc$ from Gaia and SDSS, \citet{2018MNRAS.478..611B} found that the stellar halo’s velocity ellipsoid was stretched dramatically for stars with $\rm [Fe/H]>-1.7$. They suggested that this property was caused by a major accretion event of a satellite with virial mass $M_{\rm vir}>10^{10}\, M_{\odot}$, the so-called ``Gaia Sausage'', between 8 and $11\,\rm Gyr$ ago.
Meanwhile, \citet{2018Natur.563...85H} proposed that the retrograde stars in the halo and some of the low angular momentum stars could be caused by an ancient merger, ``Gaia-Enceladus''. The components of the ``Gaia-Enceladus'' are slightly prograde with high eccentricities or strongly retrograde. These two events may represent the same merger event due to their properties and we refer to this merger as the GES. 

However, we cannot calculate the eccentricity $e$, $z$-component angular momentum $L_z$, and total energy $E$ due to the lack of radial velocities, which makes it very difficult to recognize the GES members in our sample. To investigate whether eccentricity can be constrained using only proper motions and distances, we create mock data for halo-like stars, as detailed in the Appendix.
We find that most mock stars satisfying the criteria of $V_\perp<60\,\rm km\,s^{-1}$ and $r>15\,\rm kpc$ have high $e$ ($e>0.7$), largely unaffected by radial velocities. $V_\perp$ is the tangential velocity defined by $V_\perp=\sqrt{V_l^2+V_b^2}$. This phenomenon can be explained by the following: if one star has a large $r$, the tangential velocities $V_\perp$ relative to the Sun can approximate those relative to the Galactic center. In this case, low $V_\perp$ means low rotation velocities and consequently, high orbital eccentricities. Further details and the results of this test are available in the Appendix. This approach of recognizing stars with high eccentricities just from their proper motions and distances could be useful for the large datasets lacking radial velocities.

We select groups with more than 60\% members satisfying the criteria:
\begin{equation}
    (V_\perp<60\,\rm km\,s^{-1}) ;\; (r>15\,\rm kpc)\,,
\end{equation}
leading to the identification of 85 groups (6,584 RRLs) consistent with the GES.
The distance and metallicity distributions of these stars are shown in Fig.~\ref{fig:d_FeH_HAC_VOD_GES}. We find the GES members spread over a large range of distances, from 15 to $41\,\rm kpc$, and the mean and standard deviation of the metallicities are $-1.62$ and $0.32\,\rm dex$, respectively, which are consistent with the metallicity of the GES in \citet{2018MNRAS.478..611B}, \citet{2018Natur.563...85H} and \citet{2023MNRAS.tmp.3593G}. 
%Besides, as shown in Fig~\ref{fig:lbXZV_HAC_VOD_GES}, most GES members locate in the range of $250^\circ<l<340^\circ$ and $10^\circ<b<75^\circ$, and we infer that the distant GES ($r>15\,\rm kpc$) is dense in this region. To validate this discovery, we aim to distinguish whether it is due to selection bias. First of all, the selection criteria of the GES members only constrain the  Galactocentric distance and velocities, so they have almost no restrictions on sky coverage. Then, due to the Gaia scanning law, the completeness of Gaia RRLs around the ecliptic plane is far lower than in other regions. In order to exclude the influence of the inhomogeneous sky coverage, we only consider the GES members in the PS1 RRab catalogue because the PS1 RRab stars have a high and uniform completeness down to $G\sim18$ (Mateu et al. 2020). the PS1 RRab stars belonging to GES are also dense in this region. These validations suggest that this overdensity is not caused by selection bias. However, we could not rule out the impact of our approach. Our approach just clusters the stars with similar proper motions due to the lack of radial velocity. 
Similar to the HAC, the stars of GES are very diffuse in the position and velocity space, so many stars cannot be linked together in the absence of radial velocity. 

\begin{figure*}
	\centering
	\includegraphics[width = \linewidth]{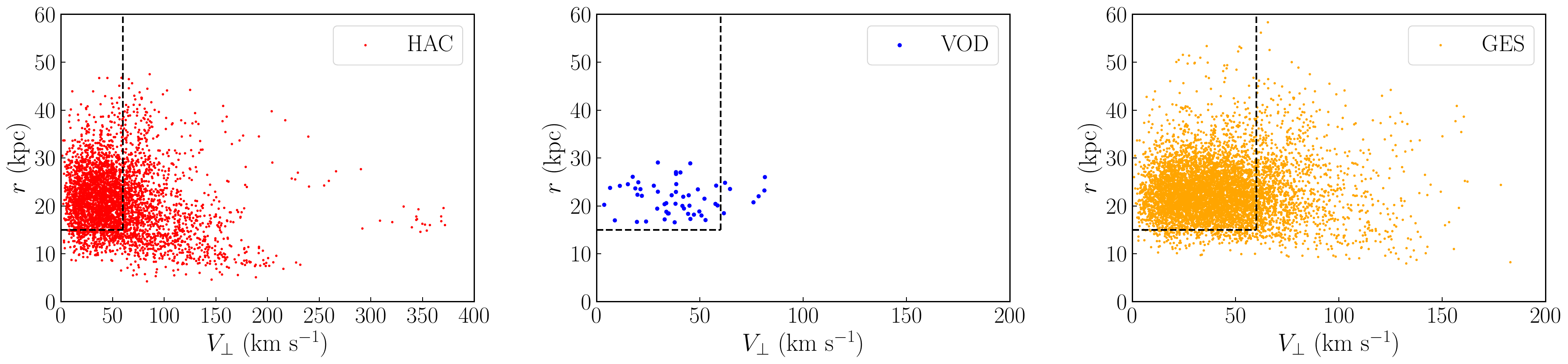}
	\caption{The distributions of the members of HAC, VOD and GES in the $(V_\perp,r)$ space. The dashed lines present the region of $V_\perp<60\,\rm km\,s^{-1}$ and $r>15\,\rm kpc$.}
	\centering
	\label{fig:Fig_Vr_HAC_VOD_GES}
\end{figure*}

\subsubsection{The Relation between GES, HAC, and VOD}
In recent years, several works \citep[e.g.][]{2016ApJ...817..135L,2021A&A...654A..15B,2022MNRAS.513.1958W,2024arXiv240703713Y} showed a link between HAC, VOD and GES. Most stars of HAC and VOD have high eccentricities and nearly zero $L_z$ angular momentum \citep{2019MNRAS.482..921S,2019ApJ...886...76D,2022MNRAS.513.1958W,2023A&A...674A..78Y,2024arXiv240703713Y}. \citet{2019MNRAS.482..921S} suggested that these two overdensities were part of the same accretion event using $\sim$350 RR Lyrae with radial velocities and Gaia DR2 proper motions. \citet{2022MNRAS.513.1958W} found similar kinematic properties between HAC, VOD, and GES, based on $\sim 3,000$ RRab stars with 6D information. Using the RRab sample from \citet{2023ApJ...944...88L}, \citet{2024arXiv240703713Y} finds that both HAC and VOD are dominated by the GES debris stars with weights of $0.67^{+0.09}_{-0.07}$ and $0.57^{+0.07}_{-0.06}$ by applying the Gaussian mixture model. As shown in these works, the orbit properties of HAC and VOD are very similar to those of GES. Our study further explores the relationship between these three substructures using a large RR Lyrae sample, which can be directly seen in Table~\ref{tab:overlap}. 12 of the 42 groups of HAC overlap with the GES according to our selection processure. Both the two groups of VOD are also identified as GES candidates. The overlap of HAC, VOD and GES can be directly observed in Fig.~\ref{fig:all_known_lbXZ}. Most members of HAC and all VOD members in the left two panels of Fig.~\ref{fig:all_known_lbXZ} have their GES counterparts in the right two panels. As shown in Fig.~\ref{fig:d_FeH_HAC_VOD_GES}, the metallicity distributions of these three substructures are remarkably similar. The distributions in the $(V_\perp,r)$ space are shown in Fig.~\ref{fig:Fig_Vr_HAC_VOD_GES}. As mentioned in Section~\ref{subsec:GES}, most stars with $V_\perp<60\,\rm km\,s^{-1}$ and $r>15\,\rm kpc$ have high $e$ and we find most of the stars ($\sim$57\% for HAC and $\sim$87\% for VOD) in HAC and VOD satisfy these criteria, suggesting that most stars in HAC and VOD have high $e$. These properties imply that the HAC and VOD may have similar origins to GES and originate from accretion events, which are consistent with the results of \citet{2019MNRAS.482..921S} and \citet{2022MNRAS.513.1958W}. 

\begin{deluxetable*}{ccc}\label{tab:overlap}
\tabletypesize{\footnotesize}
%\tablenum{3}
\tablecaption{The number of overlapping groups and RRLs between each pair of substructures. In parentheses are the number of groups and the total number of members for each substructure. The substructure pairs not listed in this table have no overlapping members.}
\tablewidth{0pt}
\tablehead{
Substructures & Overlapping groups  & Overlapping RRLs \\  }
\startdata
HAC (42 groups, 4049 RRLs) and GES (85 groups, 6584 RRLs) & 12 & 3366 \\
VOD (2 groups, 55 RRLs) and GES (85 groups, 6584 RRLs) & 2 & 55 \\
Sequoia (26 groups, 793 RRLs) and GES (85 groups, 6584 RRLs) & 2 & 73 \\
Wukong (2 groups, 65 RRLs) and HAC (42 groups, 4049 RRLs) & 1 & 48 \\
Sequoia (26 groups, 793 RRLs) and HAC (42 groups, 4049 RRLs) & 5 & 125 \\
Helmi streams (4 groups, 90 RRLs) and HAC (42 groups, 4049 RRLs) & 2 & 37 \\
\enddata
\end{deluxetable*}

\subsubsection{Attributing Groups to Helmi Streams}
\begin{figure}
	\centering
	\includegraphics[width = \linewidth]{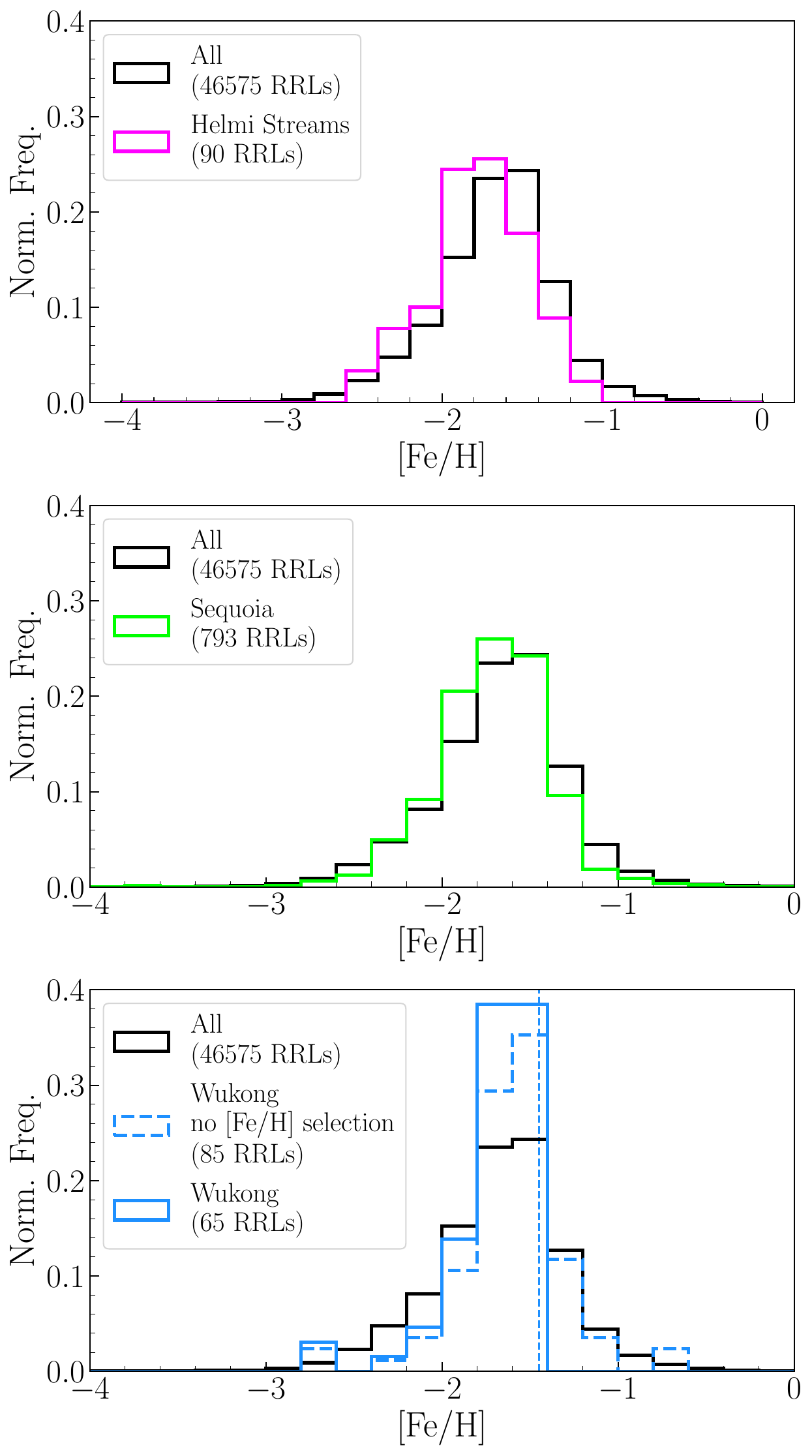}
	\caption{The metallicity distributions of the Helmi streams (top panel), Sequoia (middle panel) and Wukong (bottom panel), shown with magenta, lime and blue lines, respectively. The black lines in all panels represent the total sample. The vertical dashed line in the bottom panel represents [Fe/H]=$-$1.45.}
	\centering
	\label{fig:Helmi_Sequoia_Wukong_FeH}
\end{figure}

The Helmi streams were the first authentic inner halo kinematic substructures discovered in the solar neighborhood \citep{1999Natur.402...53H}. \citet{2019A&A...625A...5K} found nearly 600 new stars of the Helmi streams with Gaia DR2 \citep{2016A&A...595A...1G,2018A&A...616A...1G} up to a distance of 5 kpc from the Sun. The metallicity distribution in \citet{2019A&A...625A...5K} was from [Fe/H]$\sim-$2.3 to $-$1.0, with a peak at [Fe/H]$\sim-$1.5. 
We identify four groups (90 RRLs) associated with the Helmi Streams according to the selection criteria in Table~\ref{tab:diffuse_sub} (see Fig.~\ref{fig:Helmi_Sequoia_Wukong_FeH}). The mean and standard deviation of the metallicities are $-1.78$ and 0.30 dex, respectively, consistent with \citet{2023MNRAS.tmp.3593G}. Two of the four groups of Helmi streams overlap with HAC (see Table~\ref{tab:overlap}). According to Appendix, most stars not meeting the criteria $V_\perp<60\,\rm km\,s^{-1}$ and $r>15\,\rm kpc$ have eccentricity less than 0.5. No star in these two groups satisfies $V_\perp<60\,\rm km\,s^{-1}$ and $r>15\,\rm kpc$, suggesting that Helmi streams might be the composition of low-eccentricity stars in HAC, which is consistent with \citet{2024arXiv240703713Y}. 
\begin{figure}
 \centering
 \includegraphics[width = \linewidth]{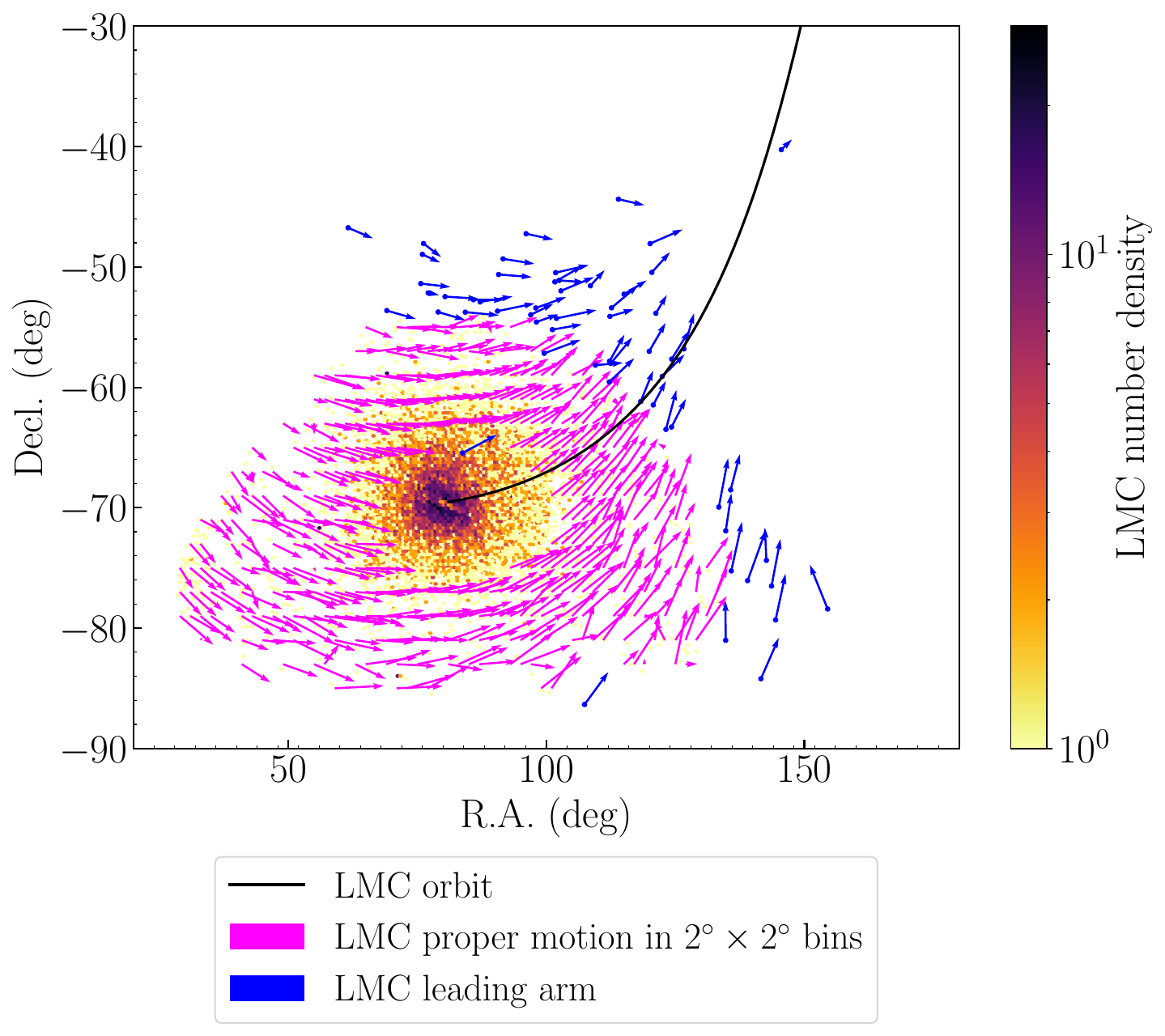}
 \caption{The sky distributions and proper motions of LMC leading arm and LMC member stars. The colorbar represents the number density of LMC members selected following Sec.~\ref{subsec:RRLsample}. Magenta arrows indicate the direction and magnitude of the proper motions of LMC members that are at an angular distance greater than 7$^{\circ}$ from the center of the LMC in $2^{\circ}\times2^{\circ}$ bins. Blue dots show the sky positions of the members of LMC leading arm with blue arrows showing their proper motions. The black line represents future orbit of LMC.}
 \centering
 \label{fig:LMC}
\end{figure}

\subsubsection{Attributing Groups to Sequoia}
Identified by \citet{2019MNRAS.488.1235M}, the Sequoia Event dominate the high-energy retrograde halo. Sequoia is clearly separated in energy and action space from the GES, which has close to zero net angular momentum \citep{2019MNRAS.488.1235M}. On average, the metallicity of Sequoia stars is 0.3 dex lower than that of GES stars \citep{2019MNRAS.488.1235M}. \citet{2020ApJ...901...48N} found another two substructures, Arjuna and I'itoi, which share the same region in $(L_z,E)$ space with Sequoia but demonstrate significantly different metallicity distributions: Arjuna with [Fe/H]$>-1.5$, Sequoia with $-2<$[Fe/H]$<-1.5$ and I'itoi with [Fe/H]$<-2$. We adopt the selection criteria in $(L_z,E)$ space in Table~\ref{tab:diffuse_sub}, in which the circularity $\eta=L_{\mathrm{z}} /\left|L_{\mathrm{z}, \max }\left(E\right)\right|$. \footnote{$L_{\mathrm{z}, \max }\left(E\right)$ is the maximum $L_z$ achievable for an orbit of energy $E$, which is computed by the angular momentum of a circular
orbit of energy $E$. Thus $\eta$ takes extreme values
+1 or $-1$ for co-planar circular prograde or retrograde orbits,
respectively. \citep{2019A&A...630L...4M}} 26 groups (793 RRLs) are identified to be associated with Arjuna+Sequoia+I'itoi according to \citet{2020ApJ...901...48N}, but only the metallicity distribution peak of Sequoia is observed in Fig.~\ref{fig:Helmi_Sequoia_Wukong_FeH}. The mean and standard deviation of the Sequoia candidates metallicities are
$-$1.72 and 0.32 dex, respectively, consistent with \citet{2019MNRAS.488.1235M}, \citet{2022MNRAS.513.1958W} and \citet{2023MNRAS.tmp.3593G}. Several works have indicated the potential relation between Sequoia and GES \citep{2019A&A...631L...9K,2020ARA&A..58..205H,2020ApJ...901...48N,2023MNRAS.520.5671H}. Our results in Table~\ref{tab:overlap} show that two of the 26 groups of Sequoia overlap with GES, in which 70\% fall in the region of $V_\perp<60\,\rm km\,s^{-1}$ and $r>15\,\rm kpc$, suggesting that Sequoia and GES have similar orbital properties. There are also five groups of Sequoia overlapping with HAC, with only 4\% satisfying $V_\perp<60\,\rm km\,s^{-1}$ and $r>15\,\rm kpc$. This part of Sequoia probably contribute to the low-eccentricity tail of HAC, too \citep{2024arXiv240703713Y}.

\subsubsection{Attributing Groups to Wukong}
Identified by \citet{2020ApJ...901...48N}\footnote{Wukong was independently reported in \citet{2020ApJ...898L..37Y} as a ``low-mass stellar debris stream 1'', LMS-1.}, Wukong resides at the prograde margin of the GES, associating with at least two globular clusters, NGC 5024 and NGC 5053 \citep{2020ApJ...901...48N,2020ApJ...898L..37Y,2021ApJ...920...51M}. By experiment, we combine the selecting criteria of Wukong in \citet{2020ApJ...901...48N} and \citet{2023arXiv230813702L} in Table~\ref{tab:diffuse_sub}, in which the actions $\mathbf{J}=\left(J_R, J_\phi, J_z\right)$ 
are calculated according to 6D kinematic information including the prior of radial velocity using the \texttt{AGAMA} library \citep{2019MNRAS.482.1525V} under the Galactic model potential of \citet{2017MNRAS.465...76M}. $J_R$, $J_\phi$ and $J_z$ are the radial, azimuthal, and vertical components, respectively, in a cylindrical frame. Two groups (85 RRLs) are identified to be associated with Wukong, shown as the dashed histogram in 
the bottom of Fig~\ref{fig:Helmi_Sequoia_Wukong_FeH}. Without chemical selection, the metallicity distribution presents a relatively metal-rich tail that likely corresponds to GES. After applying the criteria [Fe/H]$<-1.45$ in \citet{2020ApJ...901...48N}, 65 RRLs remain in these two groups, shown as solid histogram in the bottom of Fig.~\ref{fig:Helmi_Sequoia_Wukong_FeH}. The mean and standard deviation of the metallicities are $-1.71$ and 0.23 dex, respectively, consistent with \citet{2020ApJ...901...48N}. The larger one of the two groups of Wukong overlaps with HAC (see Table~\ref{tab:overlap}), none of which satisfies $V_\perp<60\,\rm km\,s^{-1}$ and $r>15\,\rm kpc$, likely corresponding to the low eccentricity component of HAC \citep{2024arXiv240703713Y}.

\begin{figure*}
	\centering
	\includegraphics[width = \linewidth]{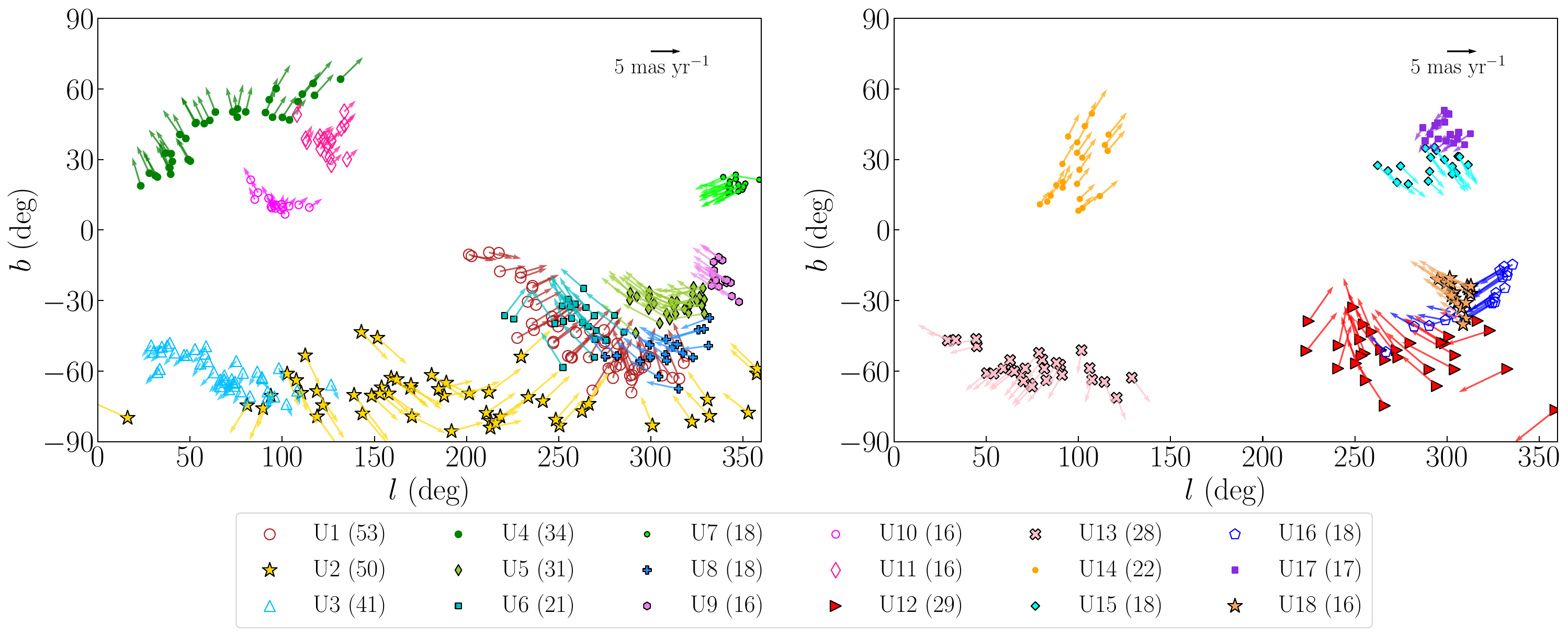}
	\caption{The spatial distributions $(l,b)$ and proper motions $(\mu_l^{*},\mu_b)$ of the 18 unknown groups named from U1 to U18. The length of the proper motion arrows are scaled by square root for clarity. The black arrows in the upper right of each panel represent the proper motion of 5$\,\text{mas}\,\text{yr}^{-1}$. }
	\centering
	\label{fig:18unknown_lbpm}
\end{figure*}

\begin{figure*}
	\centering
	\includegraphics[width = \linewidth]{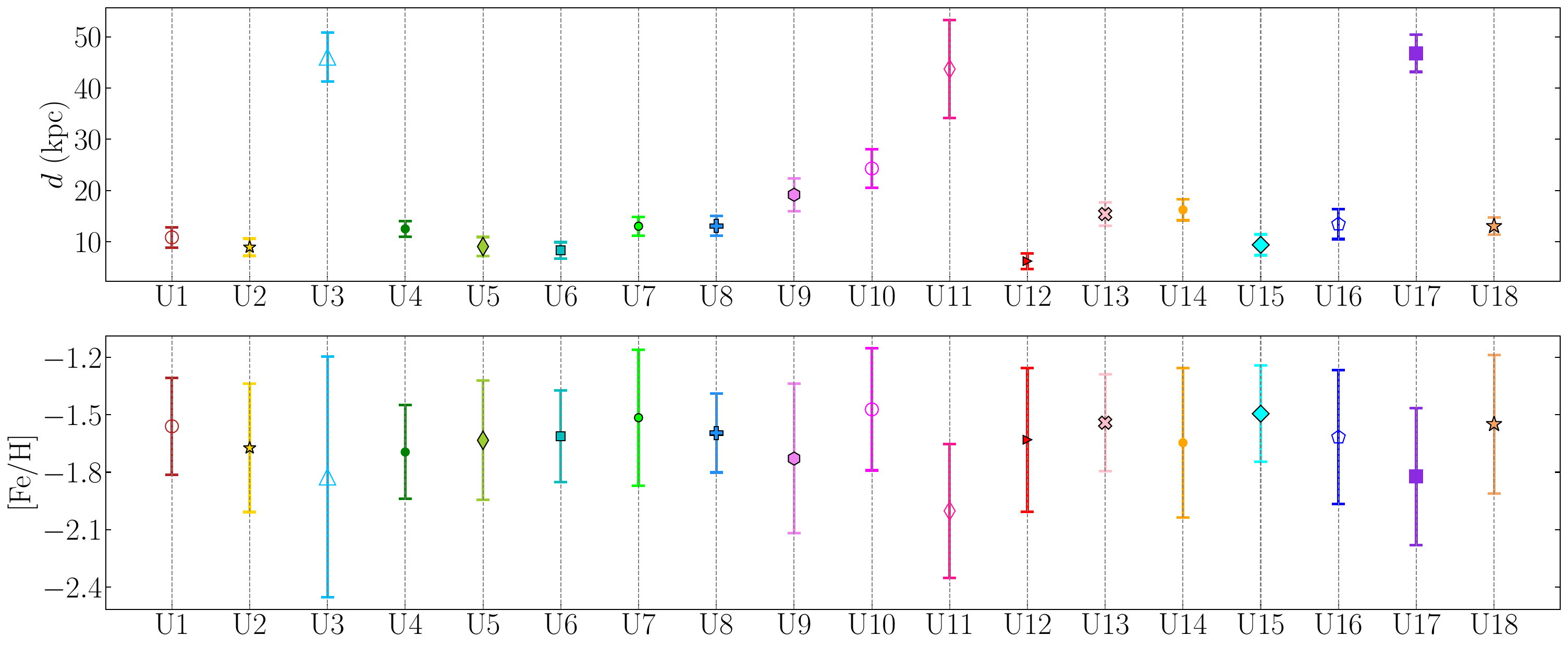}
	\caption{The distributions of heliocentric distances and metallicities of the 18 unknown groups. The symbols and errorbars represent the mean values and standard deviations, respectively.}
	\centering
	\label{fig:18unknown_dFeH}
\end{figure*}

\subsubsection{LMC leading arm}
After comparison with \citet{Petersen2022LMC}, we identified a group that appears to be part of the LMC leading arm. Using a sample of Gaia EDR3 RRL stars, \citet{Petersen2022LMC} found LMC halo members at up to $\sim 30^{\circ}$ from the LMC center, as well as LMC leading arm candidates in the Galactic north, guided by a live N-body model of MW-LMC interaction. As shown in Fig.~\ref{fig:LMC}, the members of the group we identified to be associated with the LMC leading arm have similar proper motions to the outer part of LMC and are along the LMC's future orbit integrated using \texttt{galpy} \citep{bovy2015galpy}. The means and standard deviations of the distance and metallicity of this group are $45.98\pm6.66\,\rm kpc$ and $-1.92\pm0.51\,\rm dex$, respectively. This group’s 3D position and proper motions align with the LMC leading arm, though radial velocity measurements are needed for confirmation.

\subsubsection{Comparison with \citet{2023MNRAS.tmp.3593G}}
Compared to \citet{2023MNRAS.tmp.3593G} which used the same RRL catalog from \citet{2023ApJ...944...88L} as this work, we have identified all the substructures in Table 6 of \citet{2023MNRAS.tmp.3593G} except Metal-Weak Thick Disk because we have removed disk stars from the sample. \citet{2023MNRAS.tmp.3593G} selected 5,355 RRL stars with radial velocity measurements mainly from Gaia DR3. The numbers of members we identified for GES, Sequoia, and Sgr stream exceed those of the corresponding substructures in \citet{2023MNRAS.tmp.3593G} by more than tenfold, indicating the power of our method using only 5D kinematic data to extend the understanding of known substructures.

\subsection{18 unknown groups}\label{subsec:unknown}
Apart from these known substructures we have identified, there are dozens of remaining groups. After comparison with 97 known Galactic stellar streams in \texttt{galstreams 1.0.3} \citep{2023MNRAS.520.5225M} in 3D position and proper motion spaces, 18 remaining groups are identified as unknown group candidates that are not related to any known substructures in MW. 

The spatial distributions and proper motions of the 18 unknown groups are presented in Fig.~\ref{fig:18unknown_lbpm}. Fig.~\ref{fig:18unknown_dFeH} shows the mean values and standard deviations of distances and metallicities of the members of each unknown group. These groups are named from U1 to U18 and each of them has consistent 3D position and proper motion distribution. It is worth noting that U11 is as far as $40\sim50\,$kpc and metal poor with the metallicity of $-2.0\pm0.34\,$dex, which is likely associated to an old merger in the outer halo. Nevertheless, the reliability and detailed kinematic and chemical properties of the 18 unknown groups could be further confirmed by future spectroscopic observations. The parameters of the members of the 18 unknown groups are listed in Table~\ref{tab:all_sub}.

\section{Summary}\label{sec:summary}
In this work, we identify substructures of the Galactic halo using 46,575 RRLs with proper motions from Gaia DR3 and precise refined photometric metallicities and distances from \citet{2023ApJ...944...88L}. By adopting a prior of the radial velocity, we utilize an approach similar to Xue et al. (2024, in prep.) to calculate the distributions of the five IoM parameters and identify substructures by the FoF algorithm. 

Based on the Sgr stream, we validate our method by comparing the results in several ways. These validations suggest that our method could distinguish around four-fifths of the member stars in a substructure, while misidentifications could be mostly due to the lack of radial velocity information. In addition, our method is proved to be more efficient than directly identifying substructures in the position-velocity space.

A uniform and quantitative method is applied to attribute groups to known substructures. After MC simulation with radial velocity prior, groups with higher association confidences than a threshold is considered to be linked to a particular known substructure. Members of Sgr stream and Sgr core, OC stream, Cetus-Palca, HAC, VOD, GES, Helmi Streams, Sequoia and Wukong are identified.

The Sgr stream is well recognized and clustered into three groups in our results. We estimate the mean metallicities for the leading arm, trailing arm, and Sgr core as $-1.71$, $-1.65$, and $-1.60$, respectively. Furthermore, when comparing these results with theoretical models, we find that the LM10 and Vasiliev21 models are in good agreement with our Sgr RRLs in terms of spatial and proper motion distributions. The metallicity gradient along the Sgr stream detected by the RRL is reported for the first time.

Our mock data tests reveal that most stars with $V_{\perp}<60\,\rm km\,s^{-1}$ and $r>15\,\rm kpc$ exhibit large eccentricity ($e>0.7$). This suggests a way of recognizing part of the stars with high eccentricity just from their proper motions and distances to select GES candidates. 12 of the 42 groups of HAC and both of the two groups of VOD overlap with GES. The low tangential velocities and large Galactocentric distances of HAC and VOD RRLs suggest that most of these members have high $e$. The similar kinematic and chemical properties to GES imply that HAC and VOD may have similar origins to the GES. 

Additionally, we successfully identify probable members of the Helmi Streams, Sequoia and Wukong. All of these substructures exhibit metallicity distributions that are consistent with those reported in the literature. The identification of these low mass and spatially dispersed substructures further validates the reliability of our method. The Helmi Streams, Sequoia and Wukong all have groups overlap with HAC, likely contributing to the low-eccentricity component of HAC. A group is consistent with the LMC leading arm in 3D position and proper motion, pending confirmation by radial velocity measurements. 

Finally, 18 unknown groups are discovered and consistent in 5D kinematic space, demonstrating the potential of our method for uncovering Galactic substructures and waiting for further spectroscopic study. 13,624 of 46,575 (i.e. 29\%) RRLs are associated with known substructures and unknown groups. It is also indicated that our method of identifying MW substructures using 5D kinematic data is promising for bridging the substantial gap between photometric and spectroscopic data and paving the way to detect more distant MW halo substructure in the future.

\section*{Acknowledgements}
This work is supported by the National Key R\&D Program of China No. 2019YFA0405504, the National Natural Science Foundation of China (NSFC) under grant No. 11988101,  12090040, 12090044 and China Manned Space Project with No. CMS-CSST-2021-B03. L.Z. and X.-X.X. acknowledge the support from CAS Project for Young Scientists in Basic Research grant No. YSBR-062.
This work has made use of data from the European Space Agency (ESA) mission
{\it Gaia} (\url{https://www.cosmos.esa.int/gaia}), processed by the {\it Gaia}
Data Processing and Analysis Consortium (DPAC,
\url{https://www.cosmos.esa.int/web/gaia/dpac/consortium}). Funding for the DPAC
has been provided by national institutions, in particular the institutions
participating in the {\it Gaia} Multilateral Agreement.
Guoshoujing Telescope (the Large Sky Area Multi-Object Fiber Spectroscopic Telescope LAMOST) is a National Major Scientific Project built by the Chinese Academy of Sciences. Funding for the project has been provided by the National Development and Reform Commission. LAMOST is operated and managed by the National Astronomical Observatories, Chinese Academy of Sciences.

\section*{DATA AVAILABILITY}
The data underlying this article are available in the article and in its online supplementary material.

%% For this sample we use BibTeX plus aasjournals.bst to generate the
%% the bibliography. The sample631.bib file was populated from ADS. To
%% get the citations to show in the compiled file do the following:
%%
%% pdflatex sample631.tex
%% bibtext sample631
%% pdflatex sample631.tex
%% pdflatex sample631.tex

\bibliography{sample631}{}
\bibliographystyle{aasjournal}

%% This command is needed to show the entire author+affiliation list when
%% the collaboration and author truncation commands are used.  It has to
%% go at the end of the manuscript.
%\allauthors

%% Include this line if you are using the \added, \replaced, \deleted
%% commands to see a summary list of all changes at the end of the article.
%\listofchanges

\appendix
\label{sec:appendix}
\section{Mock Data Test}
\begin{figure*}
        \figurenum{A1}
	\centering
	\includegraphics[width = 12cm]{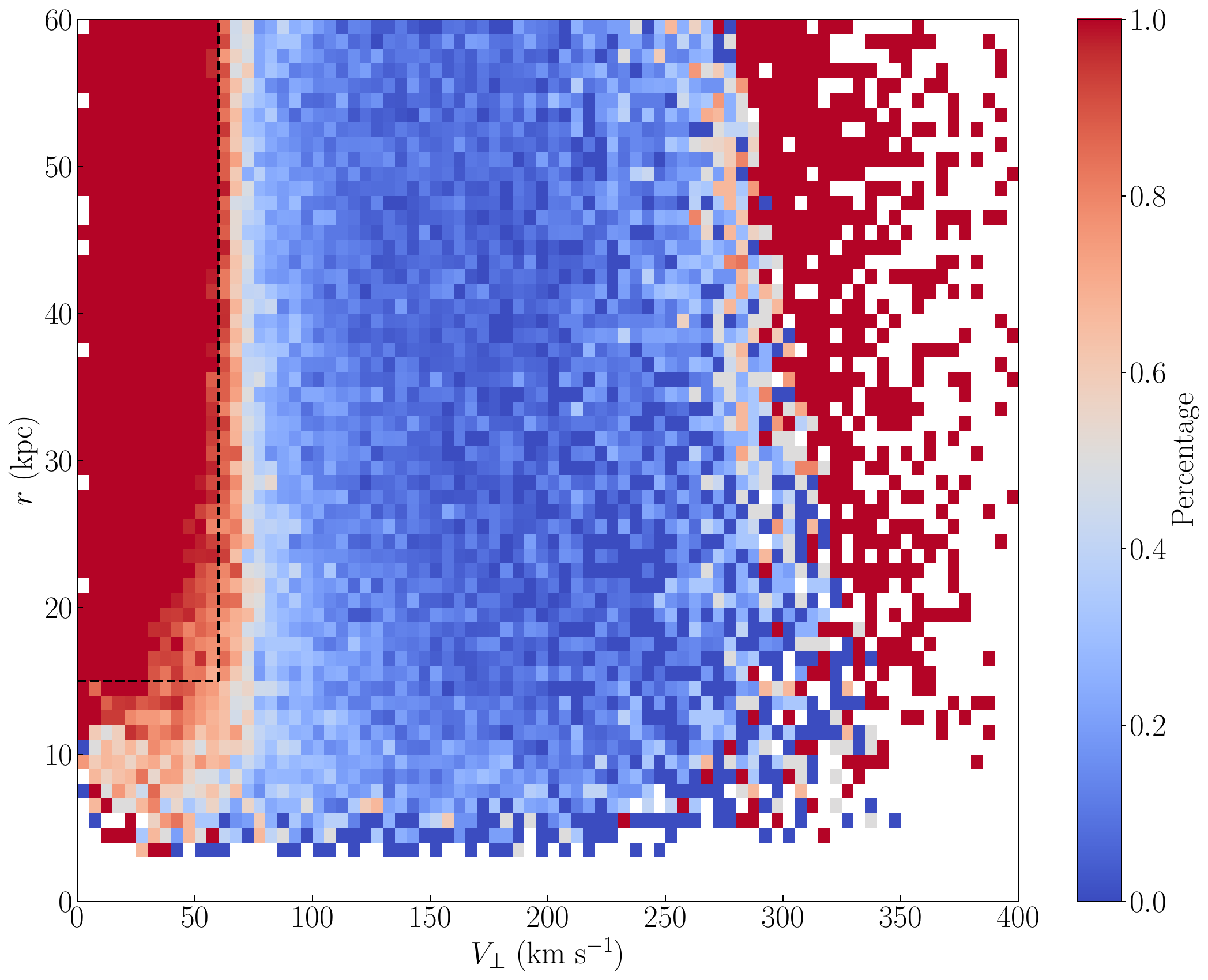}
	\caption{The map in the $(V_\perp,r)$ space of the mock data. The colour represents the percentage of the stars with high eccentricity $(e>0.7)$. The black dashed lines show the region of $V_\perp<60\,\rm km\,s^{-1}$ and $r>15\,\rm kpc$.}
	\centering
	\label{fig:A1}
\end{figure*}

\begin{figure*}
        \figurenum{A2}
	\centering
	\includegraphics[width = 12cm]{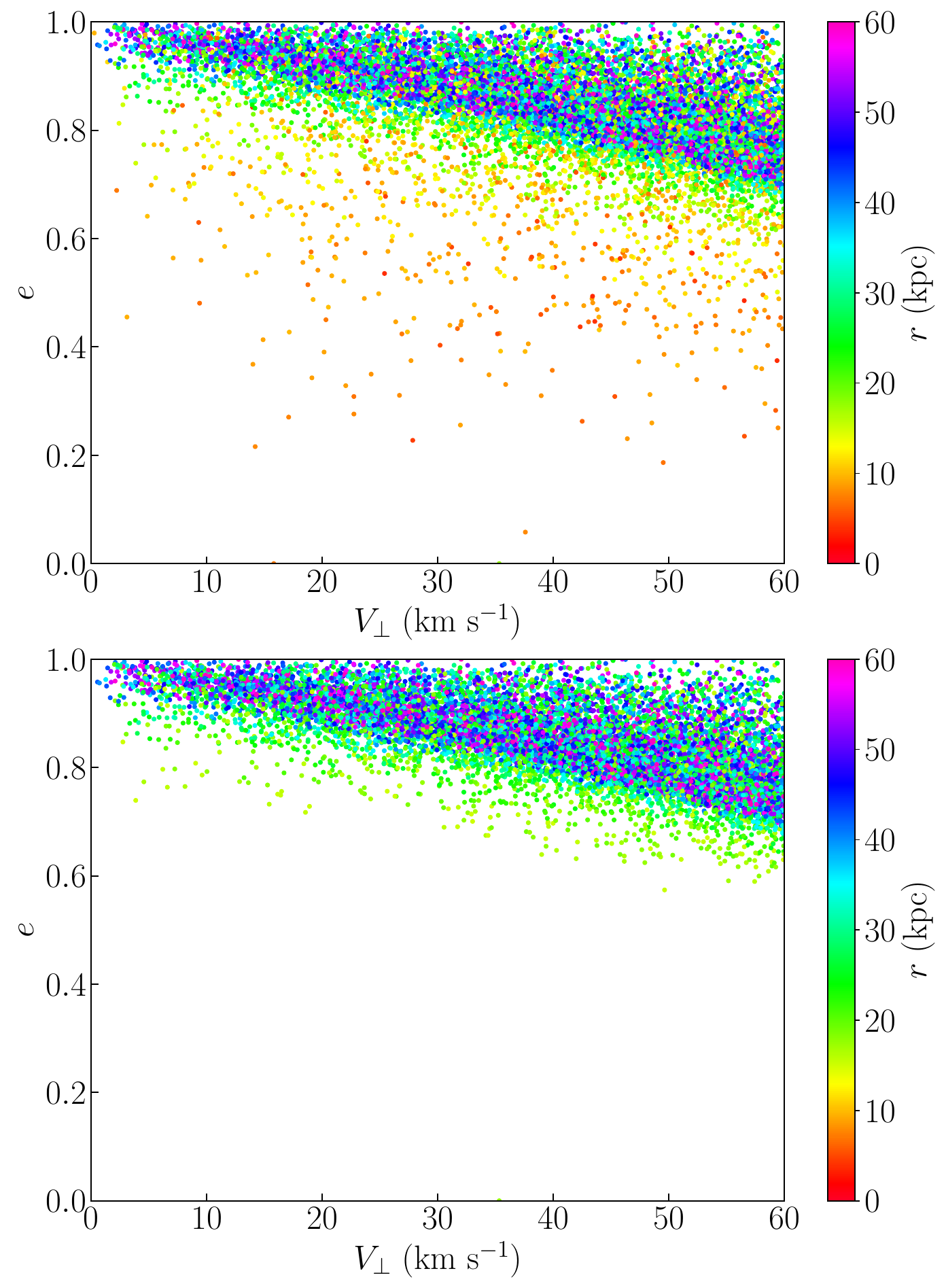}
	\caption{The distributions in the $(V_\perp,e)$ space of the mock data. The different colour represents the value of Galactocentric distance. The top panel shows all mock data with $V_\perp<60\,\rm km\,s^{-1}$. The bottom panel only shows the mock data with $V_\perp<60\,\rm km\,s^{-1}$ and $r>15\,\rm kpc$.}
	\centering
	\label{fig:A2}
\end{figure*}

\begin{figure*}
        \figurenum{A3}
	\centering
	\includegraphics[width = 12cm]{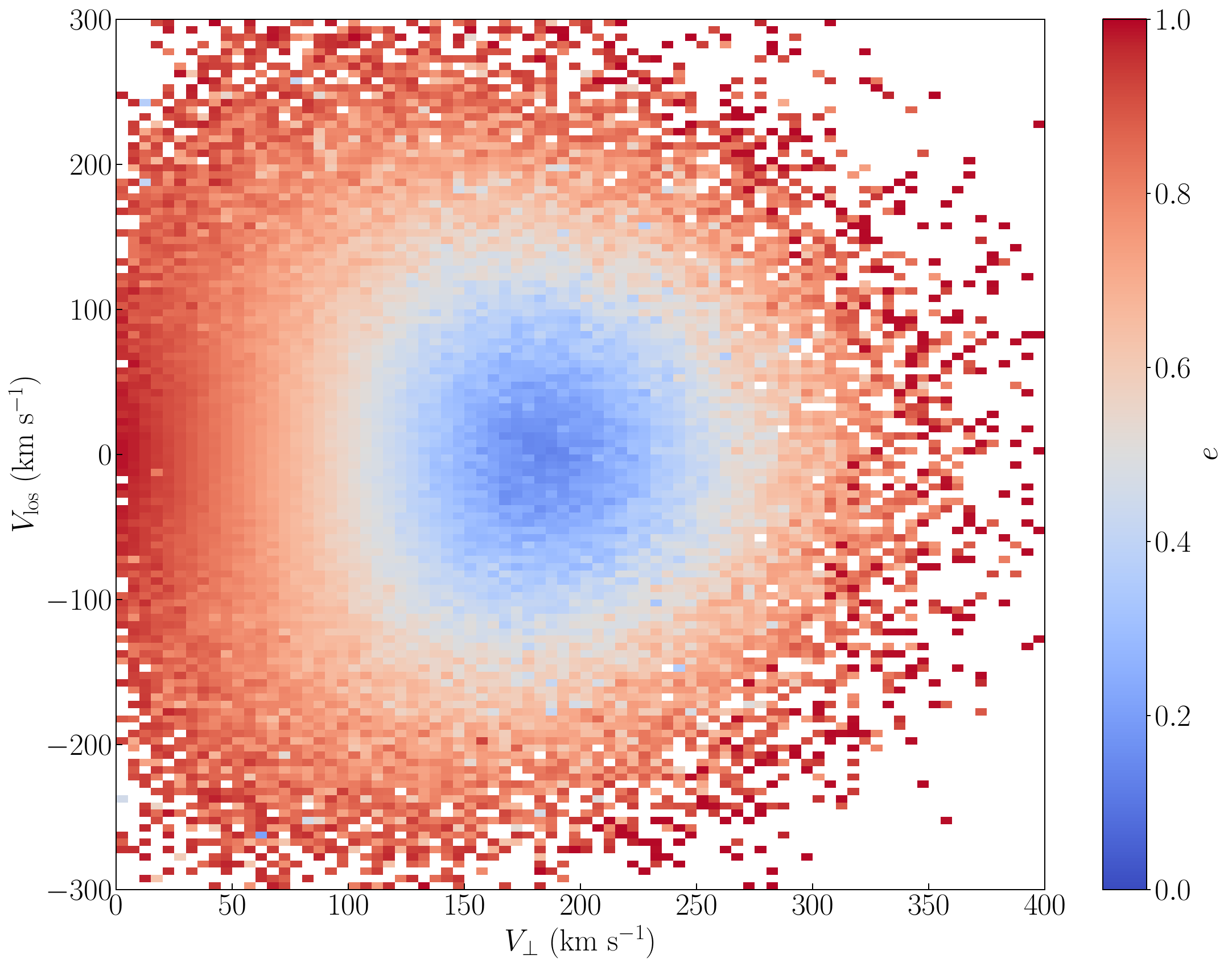}
	\caption{Density map in the $(V_\perp,V_{\rm los})$ space of the mock data with $r>15\,\rm kpc$. The colours represent the mean eccentricities.}
	\centering
	\label{fig:A3}
\end{figure*}

We use the mock data to test whether we can constrain the eccentricity when we don't have the radial velocity information in our sample. We make 100,000 halo-like mock stars. The coordinate $(l,b)$ and heliocentric distance $d$ are generated from uniform distributions with $\mathcal{U}(0,360)$, $\mathcal{U}(-90,90)$, and $\mathcal{U}(3,60)$ which represent the mock data is in all area of the sky and has the same distance range as our sample. As for the velocities, we just consider a very simple model. The $V_{\rm los}$, $V_l$ and $V_b$ in the GSR frame are generated from a same Gaussian distribution with $\mathcal{N}(0,100)$ where 0 and $100\,\rm km\,s^{-1}$ represent the mean velocity and velocity dispersion. We just consider the mock data with $|z|>3\,\rm kpc$ which is similar to the range in our sample selection. Then we calculate the eccentricity $e$ in the same Galactic potential mentioned in Section~\ref{sec:method}. Our purpose is to constrain $e$ if we just have the tangential velocities, $V_l$ and $V_b$, and the distance $d$. The percentage of stars with high eccentricity $(e>0.7)$ in the $(V_{\perp},r)$ space are shown in Fig.~\ref{fig:A1}. The $V_{\perp}$ is defined by $\sqrt{V_l^2+V_b^2}$. We find that there are two regions with a very high proportion of high $e$ stars in the relatively large $r$, one with very low $V_{\perp}$ and the other with very high $V_{\perp}$. These regions with nearly 100\% proportion of high $e$ stars imply that the influence of the radial velocities on $e$ can be neglected. We select the low $V_\perp$ region with $V_\perp<60\,\rm km\,s^{-1}$. Fig.~\ref{fig:A2} illustrates the distributions in the $(V_\perp,e)$ space of the mock data with low $V_\perp$. We find that most distant mock stars have large $e$ and the number of mock stars with lower $e$ increases as the distances become closer. We also find almost all mock samples with $V_\perp<60\,\rm km\,s^{-1}$ and $r>15\,\rm kpc$ have large $e$, which suggests that we can recognize partial high eccentricity stars by the tangential velocities and Galactocentric distance. Fig.~\ref{fig:A3} shows the distributions of the mock stars with $r>15\,\rm kpc$ in the $(V_\perp,V_{\rm los})$ space. In the low tangential velocities regions, most of these mock samples have large eccentricities regardless of the value of their radial velocities, which also suggests that the eccentricity is hardly affected by the radial velocity in the regions of $V_\perp<60\,\rm km\,s^{-1}$ and $r>15\,\rm kpc$.

\end{document}